\documentclass[twocolumn]{aastex62}

\newcommand{\tess}{\textit{TESS}}

\shorttitle{The 28-day \tess\ light curve of CD Ind}
\shortauthors{Littlefield et al.}

\begin{document}
\title{\bf Fast-Cadence \tess\ Photometry and Doppler Tomography of the Asynchronous Polar CD Ind:\\A Revised Accretion Geometry from Newly Proposed Spin and Orbital Periods}

\author{Colin Littlefield}
\affiliation{Department of Physics, University of Notre Dame, Notre Dame, IN 46556, USA}
\author{Peter Garnavich}
\affiliation{Department of Physics, University of Notre Dame, Notre Dame, IN 46556, USA}
\author{Koji Mukai}
\affiliation{CRESST and X-ray Astrophysics Laboratory, NASA/GSFC, Greenbelt, MD 20771, USA}
\affiliation{Department of Physics, University of Maryland, Baltimore County, 1000 Hilltop Circle, Baltimore, MD 21250, USA}
\author{Paul A. Mason}
\affiliation{New Mexico State University, MSC 3DA, Las Cruces, NM 88003, USA}
\affiliation{Picture Rocks Observatory, 1025 S. Solano, Suite D, Las Cruces, NM 88001, USA}

\author{Paula Szkody}
\affiliation{Department of Astronomy, University of Washington, Box 351580, Seattle, WA 98195, USA}

\author{Mark Kennedy}
\affiliation{Jodrell Bank Centre for Astrophysics, School of Physics and Astronomy, The University of Manchester, Manchester M13 9PL, UK}

\author{Gordon Myers}
\affiliation{Center for Backyard Astrophysics (San Mateo) and AAVSO, 5 Inverness Way, Hillsborough, CA 94010, USA}

\author{Robert Schwarz}
\affiliation{Leibniz-Institute for Astrophysics Potsdam (AIP), An der Sternwarte 16, 14482 Potsdam, Germany}

\correspondingauthor{Colin Littlefield}
\email{clittlef@alumni.nd.edu}

\begin{abstract}

The \tess\ spacecraft observed the asynchronous polar CD Ind at a two-minute cadence almost continuously for 28 days in 2018, covering parts of 5 consecutive cycles of the system's 7.3-day beat period. These observations provide the first uninterrupted photometry of a full spin-orbit beat cycle of an asynchronous polar. Twice per beat cycle, the accretion flow switched between magnetic poles on the white dwarf, causing the spin pulse of the white dwarf (WD) to alternate between two waveforms after each pole-switch. An analysis of the waveforms suggests that one accretion region is continuously visible when it is active, while the other region experiences lengthy self-eclipses by the white dwarf. We argue that the previously accepted periods for both the binary orbit and the WD spin have been misidentified, and while the cause of this misidentification is a subtle and easily overlooked effect, it has profound consequences for the interpretation of the system's accretion geometry and doubles the estimated time to resynchronization. Moreover, our timings of the photometric maxima do not agree with the quadratic ephemeris from Myers et al. (2017), and it is possible that the optical spin pulse might be an unreliable indicator of the white dwarf's rotation. Finally, we use Doppler tomography of archival time-resolved spectra from 2006 to study the accretion flow. While the accretion flow showed a wider azimuthal extent than is typical for synchronous polars, it was significantly less extended than in the three other asynchronous polars for which Doppler tomography has been reported.

\end{abstract}


\section{Introduction}

Although best-known for its planet-hunting capabilities, the Transiting Exoplanet Survey Satellite (TESS) promises to reap vast scientific returns across many different fields within time-domain astrophysics, particularly the study of catacylsmic variable stars (CVs). A CV is a short-period binary star system consisting of a white dwarf (WD) that accretes from a low-mass companion, usually a main-sequence red dwarf, that overfills its Roche lobe \citep{warner, hellier}. CVs with strongly magnetic WDs ($B \gtrsim 10$~MG) are known equivalently as AM Herculis stars or as polars \citep[for a review of polars specifically, see][]{cropper}. In these systems, the accretion stream from the companion star follows a ballistic trajectory until its ram pressure falls below the local magnetic pressure. The stream then flows along the WD's magnetic-field lines until it impacts the WD in an accretion region that usually emits copious X-ray bremsstrahlung radiation, as well as cyclotron radiation in the near-infrared and optical \citep{cropper}. 

The defining characteristic of polars is that the WD's rotational period equals the binary orbital period, due to the magnetic interaction between the WD and its companion. If the magnetic-field strength of the WD is too low to achieve synchronous rotation, the system is known as an intermediate polar \citep[IP;][]{patterson94}. IPs are usually highly asynchronous, with spin periods tending to be $\frac{P_{spin}}{P_{orb}}\lesssim 0.1$.

\begin{figure}
    \centering
    \includegraphics[width=\columnwidth]{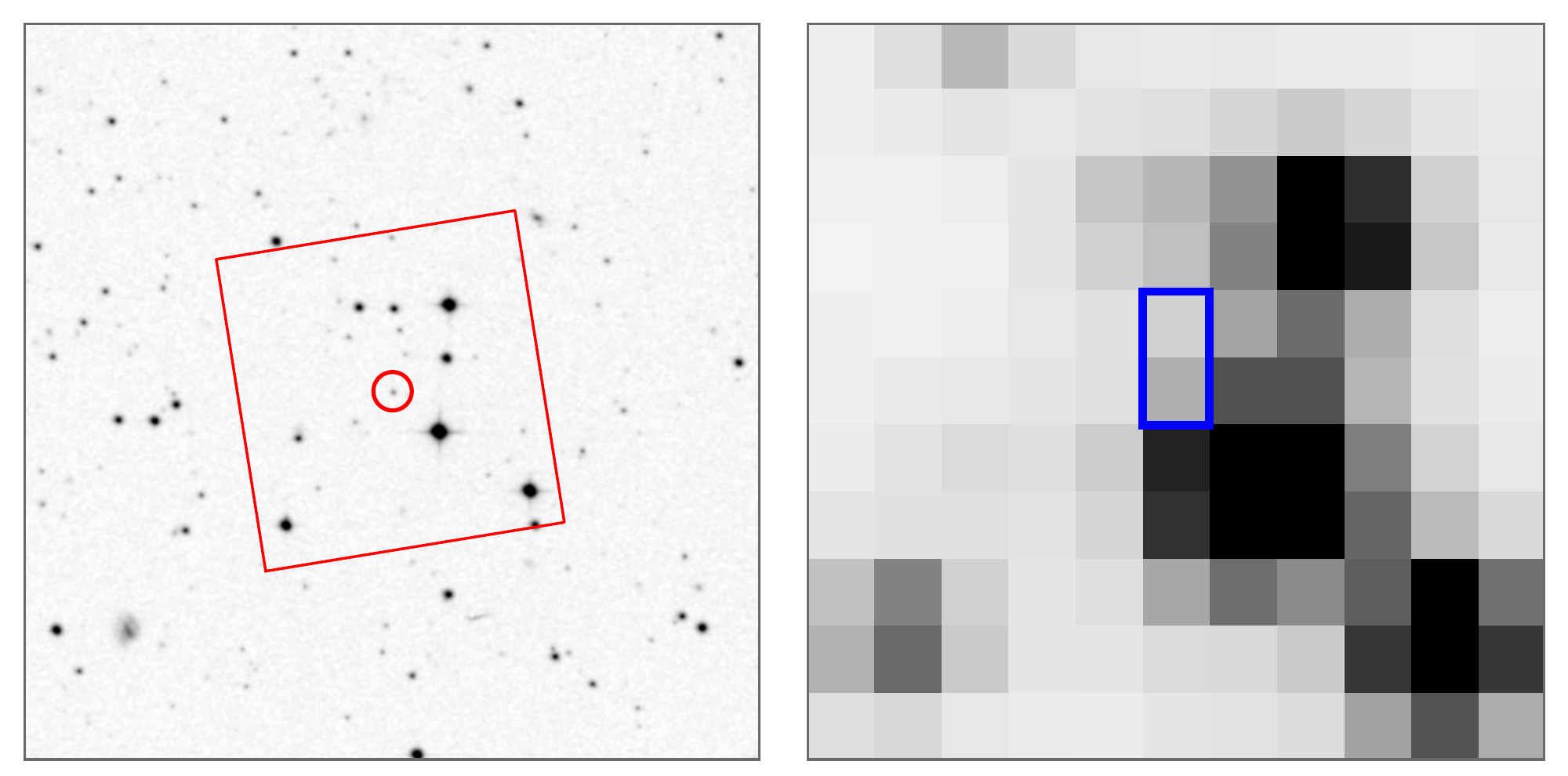}
    \caption{{\bf Left:} The \tess\ footprint (red box) superimposed on a Digitized Sky Survey (DSS) image. The small circle identifies CD Ind, and the full field of view is 8' $\times$ 8'. To match the orientation in {\tt lightkurve}, north is down, and east is left. {\bf Right:} A frame from the \tess\ target-pixel file of CD Ind. Our custom mask is outlined in blue. CD Ind is heavily blended with other sources, but using the interactive light-curve inspection functionality of {\tt lightkurve}, we identified two pixels for which the signal from CD Ind was reasonably high.}
    \label{fig:tpf}
\end{figure}

\begin{figure*}
    \centering
    \includegraphics[width=\textwidth]{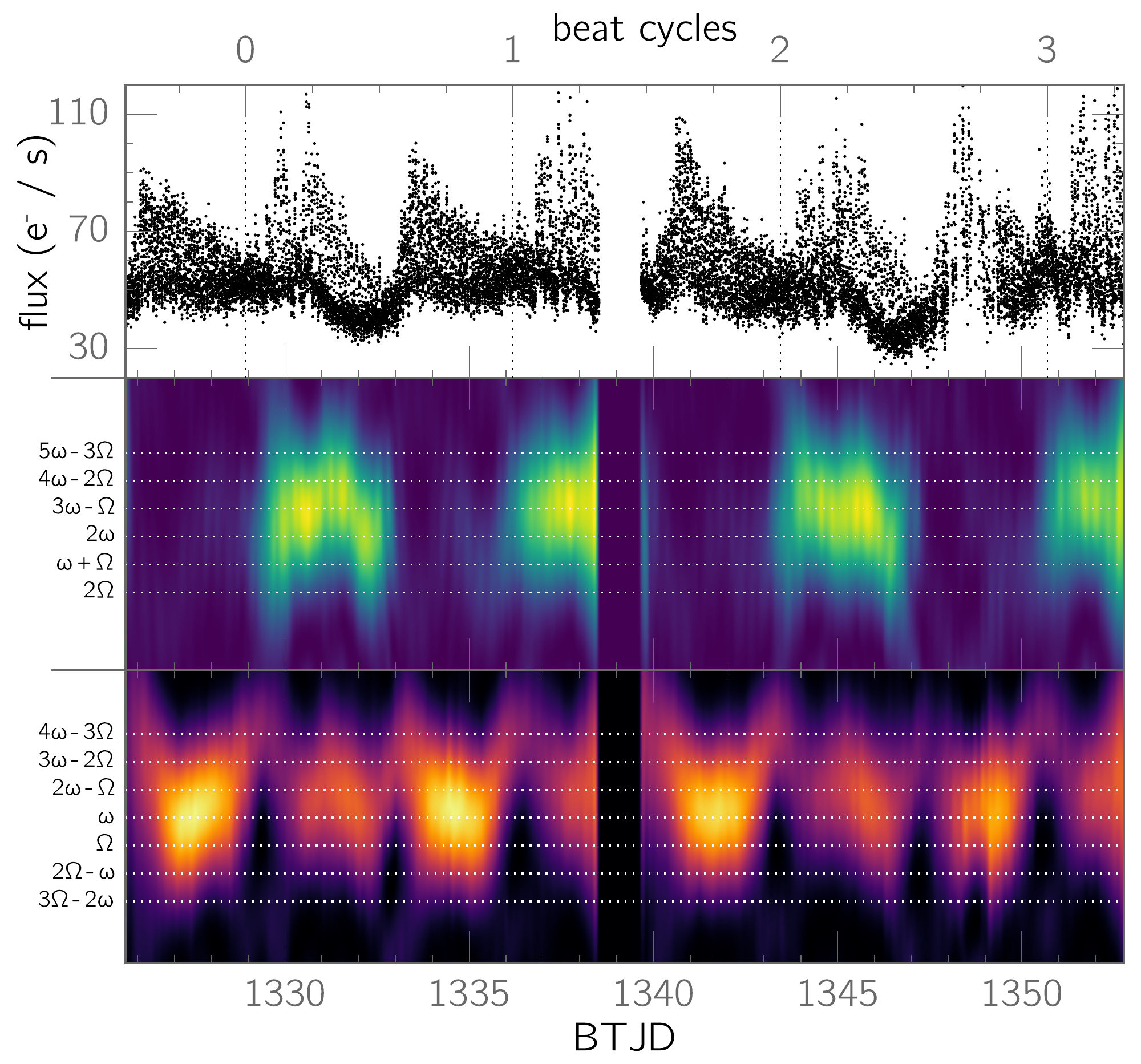}
    \caption{The full \tess\ light curve of CD Ind, along with trailed power spectra near the fundamental spin ($\omega$) and orbital($\Omega$) frequencies and their next harmonics. The trailed power spectra were constructed using a 1.5-d sliding window, and because they have different intensity cuts, we display them with different colormaps to avoid confusion. The horizontal dotted lines indicate major frequencies detected in the power spectrum of the full dataset. The frequencies and their associated periods are listed in Table~\ref{tab:frequencies}.}
    \label{fig:LC}
\end{figure*}

\begin{figure}
    \centering
    \includegraphics[width=\columnwidth]{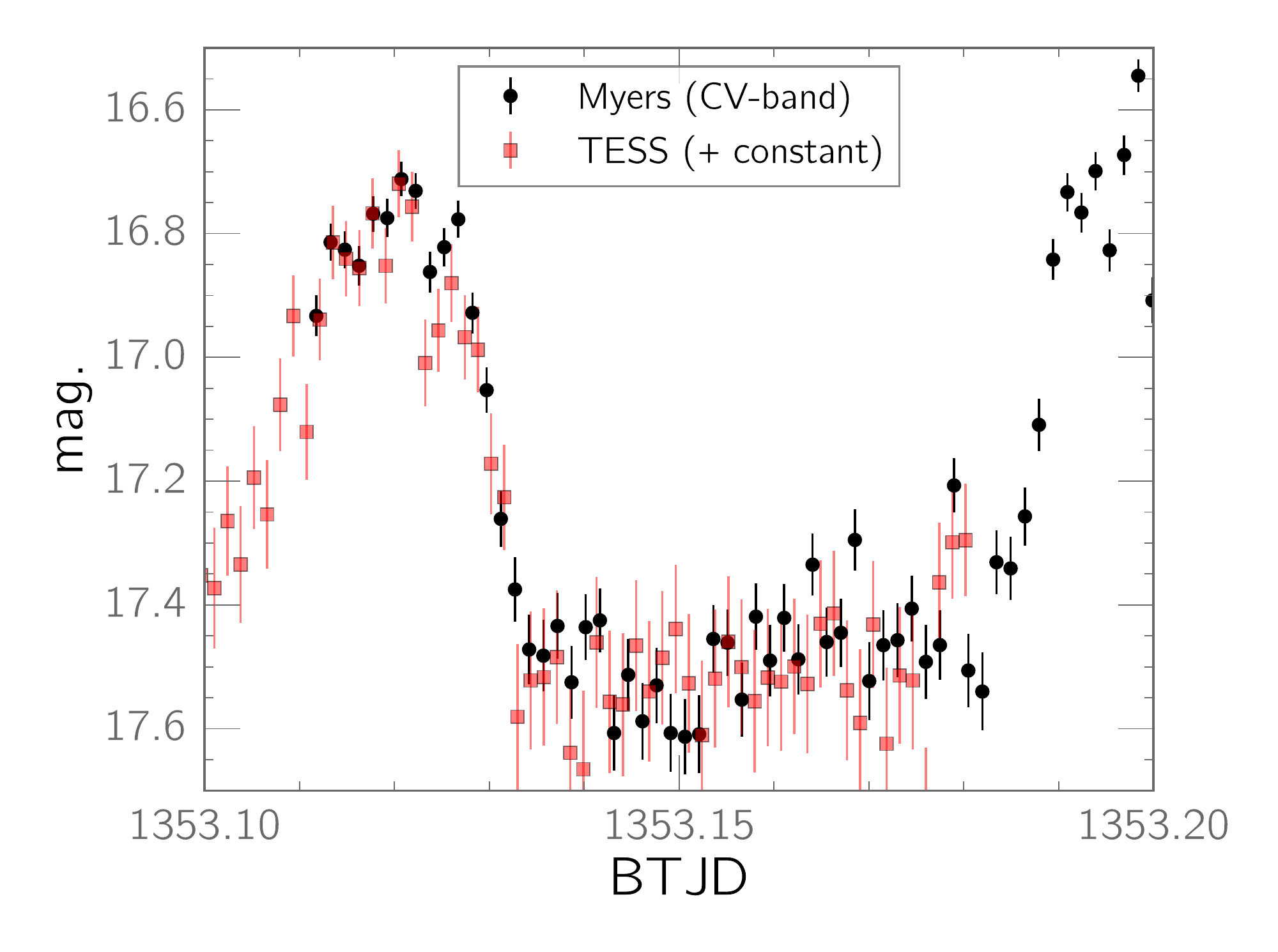}
    \caption{The overlapping light curves obtained by \tess\ and co-author Myers, who observed with an unfiltered CCD on a ground-based optical telescope. The \tess\ light curve has been converted to the magnitude scale, and we set an arbitrary zeropoint in order to vertically align the two light curves. There is no evidence of a bandpass-dependent timing offset, confirming that timing measurements of the \tess\ light curve may be reliably compared with ground-based observations at shorter wavelengths.  }
    \label{fig:overlapping_light_curve}
\end{figure}

At least six polars---including the subject of this study, CD Ind---violate the requirement of synchronous rotation by a very small ($\frac{|P_{spin}-P_{orb}|}{P_{orb}} \lesssim 3\%$) amount and are therefore known as asynchronous polars (APs).\footnote{Besides CD Ind, four other well-established APs are BY Cam \citep[][and references therein]{mason98}, V1500 Cyg \citep[][and references therein]{pavlenko}, V1432 Aql \citep[][and references therein]{littlefield}, and IGR J19552+0044 \citep{tov17}. A fifth system, 1RXS J083842.1-282723, is a very strong AP candidate according to \citet{halpern}, who identified a 44-hour beat period and a spin period that was 4\% shorter than the orbital period. However, if \citet{rea} are correct that the system's beat period is only 15~h, the spin period would be 10\% shorter than the orbital period, making it unclear whether the system would be best classified as an AP or an IP. A final system, Paloma \citep{paloma}, occupies the same twilight zone between APs and IPs; it is desynchronized by either 7\% or 14\% with respect to the orbital period, resulting in a beat period of either 0.7~d or 1.4~d.} Although both IPs and APs are asynchronous rotators, the two populations are distinguished by the magnitude of the asynchronism, its origin, and its long-term stability. IPs lack the high magnetic-field stength necessary to achieve synchronous rotation, and their spin periods are expected to evolve towards theoretically predicted equilibrium states in which the WD will remain an asynchronous rotator \citep{patterson94}. Conversely, APs are believed to be the temporary aftermath of a nova eruption in a previously synchronous polar \citep{stockman}. Each AP with a sufficiently long baseline of observations has a spin-period derivative that indicates an approach towards resynchronization, with estimated timescales between centuries and millenia \citep[e.g.,][]{myers}; thus, their asynchronous rotation is not an equilibrium state. Nevertheless, only one of the APs, V1500 Cyg, has been observed to undergo a nova, and since efforts to detect old nova shells around the other APs have yielded null results \citep{pagnotta}, the nova hypothesis has not been confirmed. Indeed, \citet{warner02} points out that if novae are the only mechanism for producing APs, the synchronization timescales estimated from observations, combined with the fraction of polars that are asynchronous, imply an unrealistically short timespan between nova eruptions in polars.

\begin{figure}
    \centering
    \includegraphics[width=\columnwidth]{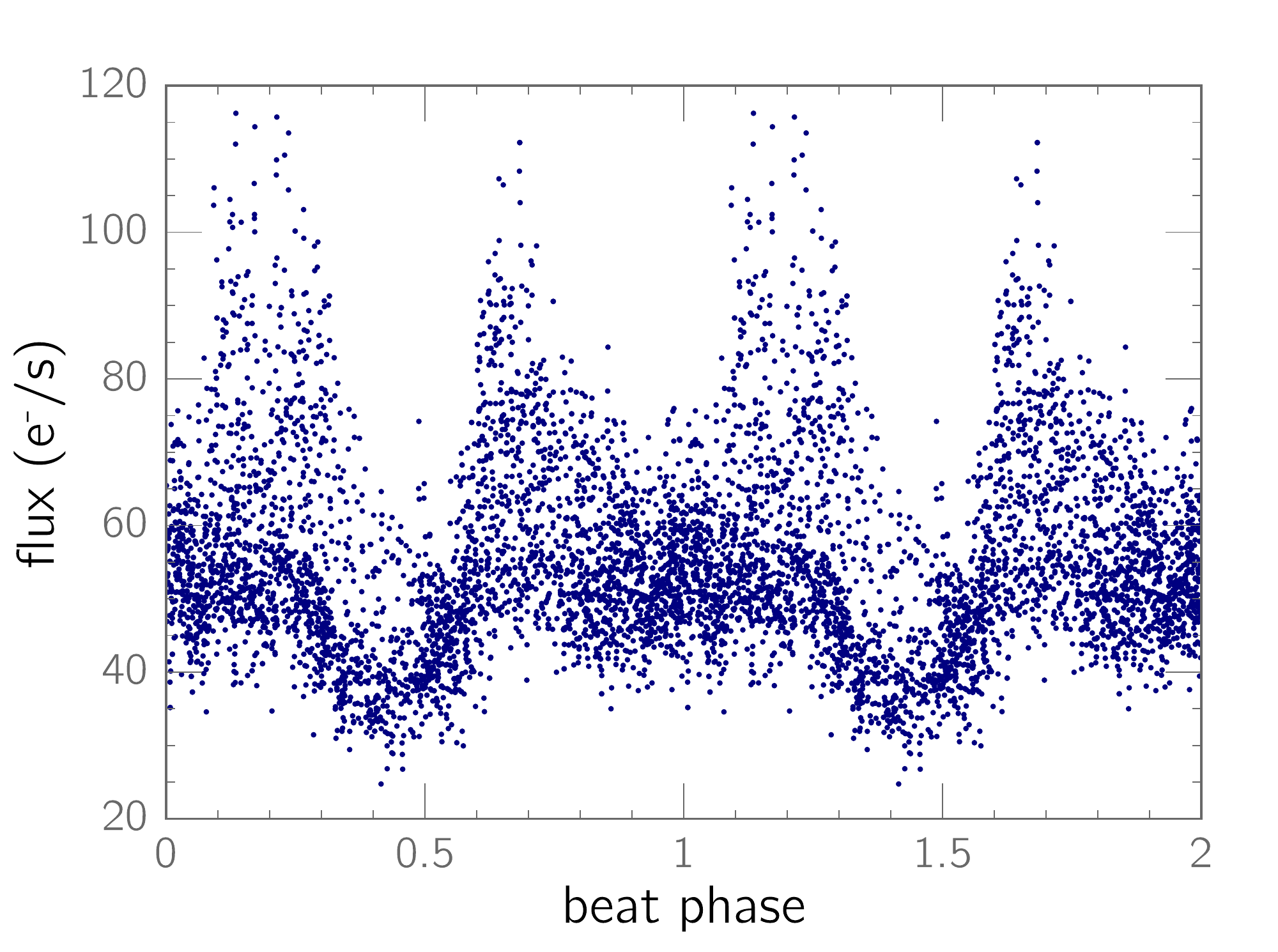}
    \caption{Waveform of the beat cycle. To improve legibility, a random subset of 20\% of the points has been plotted. As described in Sec.~\ref{sec:ephemeris}, beat phase 0.0 is defined as the beat phase at which the most pronounced pole switch occurs.}
    \label{fig:beat}
\end{figure}

\begin{figure}
    \centering
    \includegraphics[width=\columnwidth]{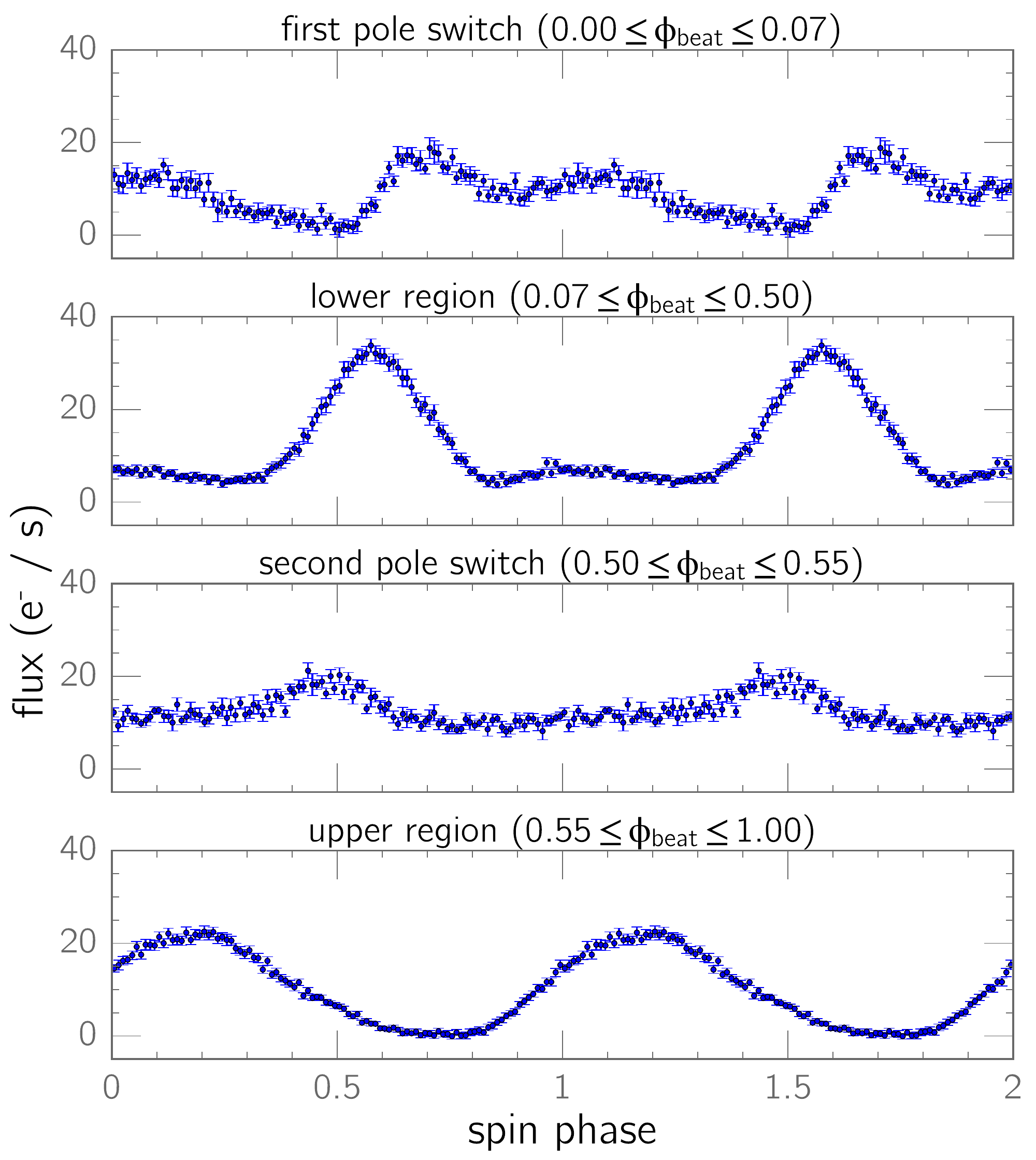}
    \caption{Waveform of the spin pulse in different sections of the beat cycle. A smoothed version of the light curve from Fig.~\ref{fig:LC} was subtracted to isolate the pulsed flux from the low-frequency variations apparent in Fig.~\ref{fig:LC}. As described in Sec.~\ref{sec:ephemeris}, beat phase 0.0 is defined as the beat phase at which the most pronounced pole switch occurs. }
    \label{fig:spin_waveform}
\end{figure}

APs provide an unmatched opportunity to study the coupling of the accretion flow to the WD's magnetic field. In synchronous polars, the accretion stream interacts with only a small subset of the WD's field lines, but in an AP, the WD slowly rotates with respect to the stream, causing the accretion flow to gradually interact with the magnetosphere across all azimuths. Viewed from an inertial frame of reference, the WD rotates at a frequency of $\omega$, but for the accretion stream and any other structure that co-rotates with the binary, the WD appears to rotate at the beat frequency between $\omega$ and the binary orbital frequency $\Omega$ ($f_{beat} = |\omega - \Omega|$). The rotation of the WD's magnetosphere with respect to the accretion flow causes the geometry of the magnetically confined portion of the flow to be highly periodic at the system's beat period ($P_{beat} = f^{-1}_{beat}$), resulting in photometric variations modulated at the beat frequency. It is even possible to detect the accretion flow switching between magnetic poles when one pole becomes more energetically favored than another.

While the beat periods in IPs are very short and are often very close to the WD spin period, APs have very long beat periods---ranging from 2.0~d for IGR J19552+0044 \citep{tov17} to 64~d for V1432 Aql \citep{littlefield}---due to the small fractional difference between $\omega$ and $\Omega$. Although it is difficult to obtain continuous observations of a full beat cycle, these observations provide snapshots of the variable accretion geometry as different field lines interact with the accretion stream. BY Cam \citep{silber, mason98}, V1432 Aql \citep{littlefield, boyd}, and CD Ind \citep{myers} have been the subjects of optical photometric campaigns that phased observations from different beat cycles. These ground-based studies have offered considerable insight into the time-averaged behavior of these systems on timescales of months-to-years, but the length of these systems' beat periods assures that any given beat cycle will be only sporadically observed, allowing short-lived phenomena to escape detection.

Although the \textit{Kepler} spacecraft observed many non-magnetic CVs, there are currently no published light curves of a polar from \textit{Kepler} data, leaving a significant void in the understanding of these objects.

\subsection{CD Ind}

CD Ind (= EUVE J2115-586, RX J2115-5840) was spectroscopically identified as a candidate polar by \citet{craig}. A follow-up spectroscopic study by \citet{vennes} verified its status as a polar and yielded estimates of its orbital period and magnetic-field strength ($B < 20$~MG). \citet{schwope} refined the field strength to $B = 11\pm2$~MG and discovered that circular polarimetry of the orbital cycle showed strong positive polarization variations on one night and negative polarization on another, consistent with the accretion flow switching between magnetic poles of opposite polarities. This observation, along with the highly variable orbital waveform of CD Ind in optical photometry, persuaded \citet{schwope} that CD Ind was most likely an AP.

\citet{ramsay99} confirmed this classification with intensive polarimetric observations of the full beat cycle. They further argued that the same ensemble of field lines accretes at all phases throughout the beat cycle, albeit with the flow traveling to a different footprint of those field lines at different beat phases. This is a remarkable conclusion because it would require the accretion flow to travel around the WD at all azimuths in order to feed that set of field lines. After finding that both accretion regions were in the same magnetic longitudinal hemisphere, \citet{ramsay99} speculated that their observations were consistent with a complex magnetic-field structure.

Around the same time, \citet{buxton} reported time-series spectroscopy obtained on five consecutive nights in 1997. They found that the spectroscopic period differed from that of \citet{vennes} by about $\sim30$~s and attributed this difference to variations in the accretion flow across the beat cycle. In addition, a phase-folded plot of their radial-velocity measurements showed clear evidence of pole-switching.

In the following year, \citet{ramsay00} reported additional polarimetry and simultaneous X-ray observations and were able to map the locations and sizes of the accretion regions across the beat cycle. Their results indicated that the lower pole accreted throughout the beat cycle, while at no beat phase did accretion occur in the immediate vicinity of the upper pole. Instead, they found evidence for an ephemeral accretion region near the magnetic equator, suggesting a complex magnetic-field structure. While their data did not yield a rigorous measurement of the orbital inclination, they determined that their polarimetry favored a relatively high orbital inclination with a small offset of the WD's magnetic axis from its rotational axis (for a dipolar field geometry).

After this initial spurt of attention in the literature, CD Ind fell under the radar and did not receive another dedicated study until \citet{myers} reported the results of a multi-year photometric campaign by the Center for Backyard Astrophysics \citep[CBA;][]{cba} between 2007-2016. They measured a spin-period derivative and used it to calculate a resynchronization timescale ($\tau = \frac{|P_{orb} - P_{spin}|}{\dot{P_{spin}}}$) of 6400 years, over an order of magnitude larger than in other APs for which $\tau$ has been measured. \citet{myers} also provide the most recent measurements of CD Ind's spin, orbital, and beat periods: 109.6564(1)~min, 110.82005(4)~min, and 7.3~d, respectively. Table~\ref{tab:frequencies} summarizes these periods.

\begin{deluxetable}{cccc}
\caption{Proposed frequency identifications \label{tab:frequencies}}
\tablehead{
\colhead{Frequency (c d$^{-1}$)} & 
\colhead{Period (min)} & 
\colhead{Original} &
\colhead{New } }
\startdata
12.86 & 112.0 & $2\Omega - \omega$ & $\Omega$ \\
13.00 & 110.8 & $\Omega$ & $\omega$ \\
13.13 & 109.6 & $\omega$ & $2\omega - \Omega$ \\
\enddata
\tablecomments{See Sec.~\ref{sec:period_identification} for an explanation of the revised frequency identifications.}
\end{deluxetable}

CD Ind has a \textit{Gaia} DR2 parallax of $4.10\pm0.10$~mas \citep{gaia, dr2}, equivalent to a distance of $243^{+5}_{-6}$~pc using the probabilistic inference procedure in \citet{BJ18}.

\section{Data}

\subsection{\tess\ light curve}

The centerpiece of our dataset is a 28-day-long light curve of CD Ind obtained at a two-minute cadence by \tess.\footnote{\citet{hakala} have independently analyzed the \tess\ observation and used particle-flow simulations to map the accretion geometry.} Since \tess\ photometry has different properties than \textit{Kepler} photometry, it is useful to briefly identify the major differences between fast-cadence data from the two spacecraft. The \textit{Kepler} bandpass extended from 400-900~nm, peaking between 500-600~nm \citep[Fig.~1 in][]{kepler_bandpass}. This is significantly bluer than the \tess\ bandpass, which covers the spectrum from 600-1100~nm. The sensitivity of \tess\ peaks near 900~nm and drops rapidly at wavelengths longer than 1000~nm \citep[Fig.~1 in][]{tess_bandpass}. Additionally, \tess's fast cadence is 2 minutes, compared to 1 minute for \textit{Kepler}. Lastly, the plate scale for \tess\ (21 arcsec px$^{-1}$) is approximately 5.3 times larger than \textit{Kepler}'s, so blending is a much greater concern for \tess\ data. However, \tess\ has much greater pointing stability than did the two-wheel \textit{Kepler K2} mission. All of these factors are relevant when comparing observations of CVs by these two spacecraft.

\tess\ observed CD Ind at a two-minute cadence with camera 2 during sector 1 between 2018 July 25 and 2018 August 22. The observations were continuous, with the exception of a 27-hour gap halfway into the sector. We used {\tt lightkurve} \citep{lightkurve} to interactively inspect the target-pixel file (TPF) and to identify a custom aperture that maximized the pulsed flux of CD Ind's light curve. Due to \tess's large plate scale, CD Ind is heavily blended with another field star---so much so that it is very difficult to visually identify CD Ind in the TPF, which is displayed in Fig.~\ref{fig:tpf}. We decided upon a custom two-pixel aperture and extracted it with the quality mask set to ``hardest.''\footnote{This setting in {\tt lightkurve} excludes any individual two-minute integrations whose quality flags indicate possible data-quality issues.} As a safeguard, we used {\tt lightkurve} to confirm that the contaminating star was not itself variable. While a constant level of contamination will suppress CD Ind's amplitude of variability, it will not impede the timing analysis of the light curve.

\begin{figure*}
    \centering
    \includegraphics[width=\textwidth]{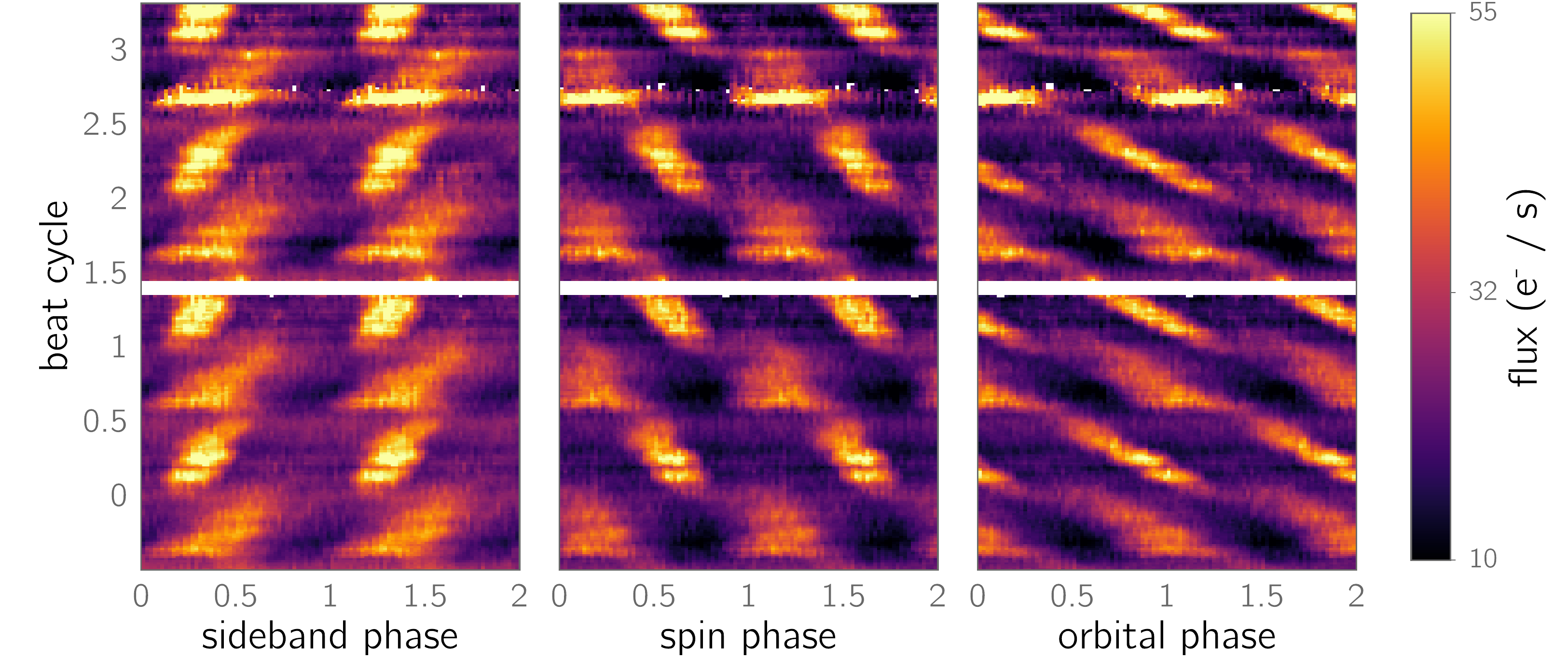}
    \caption{Two-dimensional light curves of CD Ind, phased to the re-identified $2\omega-\Omega$ sideband, spin, and orbital frequencies, using a 12-hour sliding window. The fiducial time T$_{0}$ is arbitrary for each phased light curve. The pulses are comparatively stable in phase when phased to the sideband (formerly identified as the spin frequency). The light curve phased to the newly proposed spin frequency suggests that the accretion regions are located on opposite sides of the WD. A smoothed version of the light curve from Fig.~\ref{fig:LC} was subtracted to isolate the pulsed flux from the low-frequency variations apparent in Fig.~\ref{fig:LC}. As described in Sec.~\ref{sec:ephemeris}, beat phase 0.0 is defined as the beat phase at which the most pronounced pole switch occurs.}
    \label{fig:2d_lightcurves}
\end{figure*}

\subsection{Ground-based observations}

While the \tess\ light curve is the primary focus of this study, our dataset includes ground-based photometry during the \tess\ observation from two sources: the All-Sky Automated Survey for Supernovae \citep[ASAS-SN;][]{shappee, kochanek} and co-author Myers, who observed CD Ind with an unfiltered, 50-cm telescope at a $\sim$2-minute cadence on 2018 Aug. 22. The ASAS-SN observations are too sparse for in-depth analysis, but they establish that CD Ind's $g$ magnitude varied between 16.6 and 17.4 during the \tess\ light curve, although there were also many ASAS-SN non-detections in that interval. Because the detections were near ASAS-SN's limiting magnitude, the non-detections are probably a result of the system's rapid variability causing it to briefly dip below ASAS-SN's detection threshold.

We also report time-resolved spectroscopy obtained on 2006 August 13 with the FORS1 spectrograph \citep{fors} on the Very Large Telescope Unit 1 (VLT). Lasting from 8:40 UT to 10:11 UT, the observations spanned just over 80\% of a binary orbit, consisting of 50 spectra with an exposure time of 80~s and a total cadence of 105~s per exposure. The GRIS\_1200B grism provided spectral coverage from 375~nm to 490~nm at a dispersion of $\sim$0.56 \AA\ px$^{-1}$. The slit width was 0.7 arcsec, and a degradation of the seeing during the observations resulted in significant and variable slit losses. The spectra were extracted from the CCD images using the ``twodspec.apextract" package in IRAF\footnote{IRAF is distributed by the National Optical Astronomy Observatory, which is operated by the Association of Universities for Research in Astronomy (AURA) under a cooperative agreement with the National Science Foundation.}. The pixel-to-wavelength mapping was accomplished using emission-line calibration lamps of mercury, cadmium, helium, and neon. The wavelength solution was applied and the final spectra were interpolated to be linear in wavelength using the ``dispcor" package in IRAF.

\section{The \tess\ light curve}

The top panel in Fig.~\ref{fig:LC} plots CD Ind's light curve after removal of a linear trend that caused the flux to decrease modestly over the duration of the observations, and the lower two panels show two particularly interesting frequency ranges of the trailed power spectrum. The linear trend in the light curve appears to have been instrumental in origin. When we applied a similarly shaped two-pixel measurement aperture to other stars in the TPF, most of them showed linear trends consistent with the image drifting slightly relative to the pixels. Other than that weak trend, the data quality was consistent for almost all of the light curve except between BTJD 1348-1349.5 (where BTJD = BJD - 2457000). This degradation was present in all sources in the TPF for CD Ind.

In a stroke of good fortune, co-author Myers obtained a light curve that overlapped with the final 100 minutes of the \tess\ light curve (Fig.~\ref{fig:overlapping_light_curve}), providing a means of validating the \tess\ data. The two light curves show identical variation, even down to the flickering. The observations show no measurable timing offset, and we rule out any systematic timing offset larger than 20~s.

In Fig.~\ref{fig:beat}, we plot a phased light curve of the beat cycle. Together, Fig.~\ref{fig:LC} and Fig.~\ref{fig:beat} establish the fundamental relationship between the beat cycle and the system's short-term variability. In particular, the amplitude of the fast variability and the power spectrum both underwent dramatic, periodic changes across the system's 7.3-d beat period. The beat waveform (Fig.~\ref{fig:beat}) contained two separate sections, each lasting for $\sim$45\% of the beat cycle, during which the amplitude of the short-term variability was especially high.\footnote{As we will explain in Sec.~\ref{sec:ephemeris}, beat phase 0.0 is defined as the phase of the largest discontinuity in the pulse times.} When the light curve migrated from one of these segments into the other, it went through a brief transitional period ($\sim$5\% of the beat cycle) during which the amplitude of variability dropped and the waveform became noisy. We identify these periods of chaotic variability as pole-switches, and they occur at beat phases $\sim$0.0 and $\sim$0.5. The second of these two pole-switches coincided with the end of a gentle, bowl-shaped depression that had been present in the light curve since beat phase $\sim$0.25 (Fig.~\ref{fig:beat}).

\begin{figure*}
    \centering
    \includegraphics[width=\textwidth]{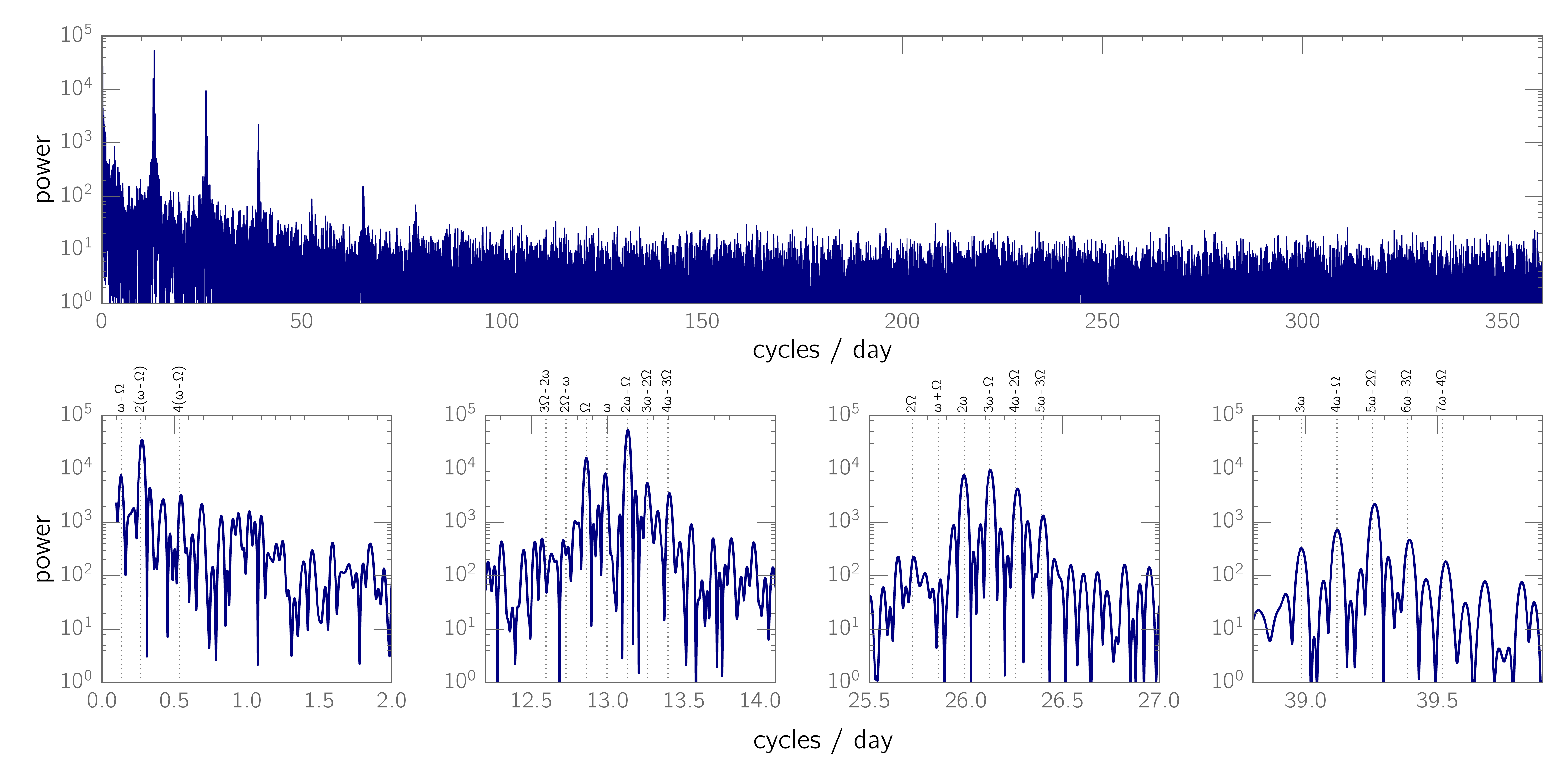}
    \caption{Lomb-Scargle power spectra of the \tess\ light curve. Top panel: the full power spectrum up to the Nyquist limit. Bottom panels: enlarged sections from the top panel showing the beat ($\omega - \Omega$), orbital ($\Omega$), and spin ($\omega$) frequencies as well as various sidebands and harmonics.}
    \label{fig:power}
\end{figure*}

\subsection{Pulse waveform}

As Sec.~\ref{sec:period_identification} will explain in detail, we use different frequency identifications than those from previous studies. Unless we expressly state to the contrary, we use only our newly proposed frequency identifications in this work.

The waveform of the 110.8-min spin pulses varied considerably across the beat cycle, and to better explore this behavior, we created two complementary visualizations. Fig.~\ref{fig:spin_waveform} separates the beat cycle into four major sections and displays the phase-averaged spin profile in each, while Fig.~\ref{fig:2d_lightcurves} consists of two-dimensional light curves phased to the newly proposed values for the spin frequency $\omega$, orbital frequency $\Omega$, and sideband frequency $2\omega-\Omega$.

As shown in Fig.~\ref{fig:spin_waveform}, the beat cycle began with a pole-switch that lasted until $\phi_{beat}\sim0.07$. The spin profile during this interval contained two weak humps separated in phase by $\sim$0.4. Such a light curve could arise from a brief epoch of simultaneous accretion onto the two poles whose individual light curves are seen in the second and fourth panels of Fig.~\ref{fig:spin_waveform}. After the pole-switch was complete, the amplitude of the pulses surged. The narrow width of these pulses is consistent with an origin in an an accretion region that rotated behind the limb of the WD for just over half of the rotational cycle; we will refer to this pole as the ``lower pole.'' The third panel in Fig.~\ref{fig:spin_waveform} establishes that the waveform during the second pole-switch was especially chaotic, and only one weak hump was discernible. During the final segment of the beat cycle, the spin waveform became quasi-sinusoidal, displaying smooth, nearly continuous variation. Because this is consistent with that accretion region being almost continuously visible, we refer to this as the ``upper pole.'' The pulses for both poles are likely shaped by the viewing-angle dependence of beamed cyclotron radiation (and in the case of the lower pole, by geometric occultation); moreover, based on our re-identification of the spin frequency, the two pulses were visible at roughly opposite spin phases.

The phased, two-dimensional light curve in Fig.~\ref{fig:2d_lightcurves} underscores that the behavior of the spin pulse was even more complicated than Fig.~\ref{fig:spin_waveform} would suggest. The spin-phased light curve shows that while the upper pole's pulses were relatively stable in phase, the pulses from the lower pole drifted significantly towards earlier spin phases for about one quarter of the beat cycle. Moreover, while the lower-pole pulses showed a relatively gradual rise to maximum amplitude after a pole switch, the upper-pole pulses appeared near their maximum amplitude immediately after a pole switch and gradually faded away.

\subsection{Power spectrum}

In the trailed power spectrum in the lower panels of Fig.~\ref{fig:LC}, the reidentified spin frequency $\omega$ dominated the second half of the beat cycle, and almost no power was present near $2\omega$ or any of the nearby sidebands. By contrast, during the first half of the beat cycle, the power near $\omega$ was less concentrated, and it is unclear which frequency was strongest; moreover, considerable power appeared near $2\omega$ and $3\omega-\Omega$ during that interval. 

The power spectrum of the full light curve (Fig.~\ref{fig:power}), which contains a rich assortment of sidebands and harmonics related to the spin frequency $\omega$ and the orbital frequency $\Omega$, provides an interesting contrast with the trailed power spectra. Although the $2\omega-\Omega$ sideband never stood out in the trailed power spectra, it was the strongest signal in the power spectrum of the full dataset. Meanwhile, the power at $\omega$ in the full power spectrum was surprisingly weak in comparison to both $2\omega-\Omega$ and $\Omega$, even though in the trailed power spectrum, it appeared to always have at least equal power compared to those frequencies. As Sec.~\ref{sec:period_identification} will explain in detail, this seemingly paradoxical behavior provides one of the foundational arguments in support of our re-identification of the spin and orbital frequencies.

Other conspicuous frequencies in Fig.~\ref{fig:power} include $2\omega$, $3\omega - \Omega$, and $5\omega - 2\Omega$. At lower frequencies, the beat frequency $\omega - \Omega$ and two harmonics were readily apparent.

\section{Identification of the Spin and Orbital Frequencies} 
\label{sec:period_identification}

\subsection{Theoretical considerations}

The starting point in the analysis of any AP is the accurate identification of the WD spin frequency $\omega$ and binary orbital frequency $\Omega$, as their inequality is the defining attribute of APs. Unfortunately, conventional methods of period identification, such as the Lomb-Scargle periodogram, can yield misleading results in the study of AP light curves. \citet{mason95, mason98} and \citet{mason96} used principles from theoretical X-ray power spectra \citep{wk92} to explain how in certain accretion geometries, the phenomenon of pole-switching can shift power from $\omega$ into both $\Omega$ and the $2\omega - \Omega$ sideband. In extreme cases, the $2\omega - \Omega$ sideband can dominate the optical power spectrum, as it does in long-term time-series photometry of BY Cam \citep{mason98}. An additional complication is that the apparent period of an AP can be biased by the length of the observations and the portion of the beat cycle that was observed \citep{mouchet}. 

One reason for this confusing behavior stems from an implicit and normally uncontroversial assumption of most period-finding algorithms: that a periodic signal does not experience internal phase jumps. The process of pole-switching in an AP can very strongly violate this assumption. In an AP in which two diametrically opposed accretion regions are each active during opposite halves of the beat cycle,\footnote{Quantitatively, this geometry is possible if the orbital inclination ($i$), magnetic colatitude ($m$), and accretion-region opening angle ($\beta$) satisfy the condition $i + b + \beta > 90^{\circ}$ \citep{mason98}.} the individual light curves of each pole would occur at opposite spin and beat phases in relation to each other. Their combined light curve would therefore show two periodic signals, each with the same frequency and visible for different halves of the beat cycle, but they would be out of phase by $\sim$half of a WD rotation.

A Lomb-Scargle periodogram will struggle in this situation, because at each trial frequency, it attempts to model the light curve with a sinusoid for which internal phase shifts are not allowed. One sinusoid at the true spin frequency $\omega$ would therefore be able to fit the light curve for either pole separately, but not both simultaneously. Thus, the algorithm has a systematic bias towards frequencies that force the two poles' pulses to remain as stable in phase as possible, even though this frequency can differ from that of the two poles' individual light curves. Most other period-finding algorithms will also yield misleading results because of similar considerations.

In an AP containing two diametrically opposite accretion regions that each accrete for different halves of the beat cycle, one frequency that satisfies this criterion is the $2\omega - \Omega$ sideband. If the light curve is phased at $2\omega - \Omega$, the spin pulses from the two poles would be forced to appear to be roughly in phase with each other, as is the case in Fig.~\ref{fig:2d_lightcurves}. If this sideband were mistakenly identified as the spin frequency, it would seem as though both regions are located in the same longitudinal hemisphere. Although this geometry is possible with a sufficiently complex magnetic field, it is important to consider on a case-by-case basis whether such a result is attributable to the combination of pole-switching and the mathematical biases of period-searching algorithms.

\subsection{A re-identification of $\omega$ and $\Omega$}

As Table~\ref{tab:frequencies} summarizes, we propose that CD Ind's true spin and orbital frequencies have long been misidentified and that the correct identifications can be obtained by adding $\omega - \Omega$ to the previous frequency identifications of the spin and orbital frequencies. For example, the former spin frequency $\omega$ becomes the newly proposed $2\omega - \Omega$ sideband. It is worth noting, however, that the beat frequency would remain the same. While it is uncomfortable to make a sweeping proposal that affects a number of previous works, an examination of the \tess\ light curve through the lens of \citet{wk92}, \citet{mason95, mason98}, and \citet{mason96} provides a compelling justification for our proposed re-identification of the spin and orbital frequencies.

The dominant short-period signal in the power spectrum of the full \tess\ light curve occurs at the long-accepted spin frequency: 13.13 cycles d$^{-1}$. If this were the correct identification, it would mean that the two next-strongest frequencies in Fig.~\ref{fig:power} would be $2\Omega - \omega$ and $\Omega$ (in order of power). This distribution of power is difficult to reconcile with the aforementioned models of pole-switching, which predict that pole-switching within a light curve will shift power away from the true rotational frequency and approximately equally into $\Omega$ and $2\omega - \Omega$ (\textit{i.e.}, $\omega\pm(\omega-\Omega)$). In particular, \citet{wk92} do not predict a significant signal at $2\Omega - \omega$. Thus, if the original frequency identifications are correct, they would pose a significant challenge to the modeling by \citet{wk92} and its subsequent extension to optical power spectra in \citet{mason95, mason98}. Additionally, the long-standing identifications do not offer a clear explanation of why the original spin frequency (our proposed $2\omega-\Omega$) dominated the full power spectrum but not the trailed power spectrum.

Instead, we identify the strongest frequency in Fig.~\ref{fig:power} as $2\omega - \Omega$, as \citet{mason95, mason98} did for a multi-year photometric dataset of BY Cam. Because the beat frequency $\omega - \Omega$ can be visually identified from the light curve, it is straightforward to identify the two next-strongest frequencies as $\Omega$ and $\omega$. After our re-identification, $\omega$ shows significantly less power than both $2\omega - \Omega$ and $\Omega$ in the full power spectrum, a distribution of power that is in better accord with the predictions of \citet{wk92} than was the distribution of power implied by the long-standing frequency identifications.

The trailed power spectrum in Fig.~\ref{fig:LC} provides additional support for our re-identifications of the frequencies. At no time did $2\omega - \Omega$ dominate the power spectra of the 1.5-day increments used to generate that figure. Instead, $\omega$ was the strongest signal when the upper pole was active, and although it is unclear which frequency was strongest when the lower was active, it certainly was not $2\omega - \Omega$. The predominance of $2\omega - \Omega$ in the power spectrum of the full dataset is consistent with the effect of pole-switching between two accretion regions on opposite sides of the WD \citep{mason95, mason98}.\footnote{In the original pre-print of this manuscript, we turned to Occam's razor to argue that an accretion geometry with two spots on opposite sides of the WD is simpler and more probable than one with accretion spots in the same longitudinal hemisphere. Our referee pointed out that the pre-polar WX~LMi, which shows two accretion spots in the same latitudinal hemisphere, provides a strong counterargument, and in hindsight, we concur. On one hand, the magnetic-field strength from a $2^{l}$ multipole component has a radial dependence of $r^{-(l + 2)}$, so at large distances from the WD, the dipole component should dominate unless the multipole component is unusually strong. However, our invocation of Occam's razor falls apart because the dipole component can be de-centered, leading to a geometry with two accretion spots in the same longitudinal hemisphere.}

Our spin-phased light curve from Fig.~\ref{fig:2d_lightcurves} would have been the orbital light curve under the long-standing identifications. While a quasi-dipolar field geometry offers a simple explanation for why the two poles would be separated in spin phase by $\sim0.5$, it is much more difficult to explain why the two regions would be visible at opposite orbital phases, as would be the case for the original frequency identifications. Furthermore, in a geometry with two accretion regions in the same magnetic longitudinal hemisphere, it is unclear to what extent one pole might become significantly more favored than the other, a prerequisite for pole-switching.

We also considered and rejected the possibility that the long-accepted spin and orbital frequencies have been confused with each other. This scenario would result in the identification of the signal at 13.13 cycles d$^{-1}$ as the orbital frequency, which is problematic because \citet{myers} identified in this signal a plausible spin-period derivative. The magnitude of this period derivative ($\dot{P} = 3.5\times10^{-10}$) is far too large to be an orbital-period derivative, so if the spin and orbital periods had been interchanged, the only explanation for the \citet{myers} period derivative would be to invoke an unstable accretion geometry that just happens to mimic the observational effects of a spin-period derivative. While not absolutely impossible, this scenario is rather contrived and unlikely.

\subsection{Reevaluating the evidence of the long-standing period identifications} \label{sec:period_reevaluation}

In light of our new identifications of $\omega$ and $\Omega$, we assess the evidence in support of their long-accepted identifications and conclude that our proposal is consistent with previous observational evidence.

The commonly accepted identification of the orbital frequency derives largely from \citet{vennes}, who found two aliases of the system's spectroscopic period, only one of which turned out to be consistent with a photometric period \citep{schwope}. \citet{vennes} measured both the base and peak components of the H$\beta$ and He II $\lambda$ 4686 \AA\ lines and found in each instance that the radial velocity variations had the same period (subject to the aforementioned alias). Based on this observation, the authors concluded that the spectroscopic period was probably the orbital period, a sound inference when observing a synchronous polar (as CD Ind was then believed to be). However, the subsequent discovery of the WD's asynchronous rotation undercuts this identification, and our spectra (Sec.~\ref{sec:spectra}) suggest that the brightest part of the flow corotates with the WD---\textit{i.e.}, at a frequency of $\omega$, not $\Omega$. Thus, the assumption that CD Ind's spectroscopic period equals its orbital period might not be sound.

Only one other published spectroscopic period is sufficiently precise to discriminate between the original spin and orbital frequencies. \citet{buxton} measured the radial velocity of the H$\beta$ line on five consecutive nights, obtaining a period of 111.3 minutes; while they did not specify the uncertainty of this measurement, the authors found that the difference between their period and its counterpart of $110.75\pm0.06$~min from \citet{vennes} was significant. The disagreement between these two measurements of the spectroscopic period would be difficult to explain if the line-emitting regions were fixed in the binary orbital orbital frame, and along those lines, \citet{buxton} concluded that changes in the accretion geometry make it difficult to accurately identify the orbital period from spectroscopy. 

Without an ironclad identification of the orbital frequency, the case for the original identification of the spin frequency becomes weaker, particularly in light of the unusual power spectra that result from pole-switching in BY Cam \citep{mason98}. The power spectra in both \citet{schwope} and \citet{ramsay99}, which together provide the basis for the long-accepted spin frequency, contained several candidate frequencies and aliases. The widely assumed equality between the spectroscopic and orbital periods provided a means of distinguishing between these signals, but without that assumption, other identifications of the frequencies are possible. For example, the top two panels in Fig.~3 of \citet{ramsay99} plot CD Ind's circular polarization as a function of the originally identified spin and orbital phases. Under our proposal, their orbitally-phased plot actually shows the spin-folded  polarimetry, and it would be generally consistent with two accretion regions of opposite polarities on opposite hemispheres of the WD.

Moreover, an inherent limitation of ground-based observations of APs is that it is almost impossible to uniformly sample the beat cycle. As \citet{mouchet} and \citet{ramsay00} noted, non-uniform sampling of a beat cycle can result in a biased period identification in an AP. The combination of an unfavorable window function and an intrinsically variable power spectrum greatly complicates power spectral analysis of ground-based observations. By contrast, the almost uninterrupted \tess\ light curve results in an essentially pristine power spectrum whose interpretation is far less susceptible to sampling effects.

We conclude, therefore, that nothing in the literature precludes our reidentification of the spin and orbital periods. The ultimate test of our proposed frequency identifications will be to measure the orbital motion of the donor star from its spectral lines, a prospect that we discuss in Sec.~\ref{sec:spectra}.

Finally, we wish to emphasize that although we disagree with the frequency identifications that emerged from the previous literature on CD Ind, the long-standing identifications were logical inferences given the limits of the available data. If the original frequencies were indeed misidentified, it would attest to the both the enigmatic nature of APs and the complex observational challenges that they pose---and not to an oversight by any of the previous studies of CD Ind.

\begin{figure}
    \centering
    \includegraphics[width=\columnwidth]{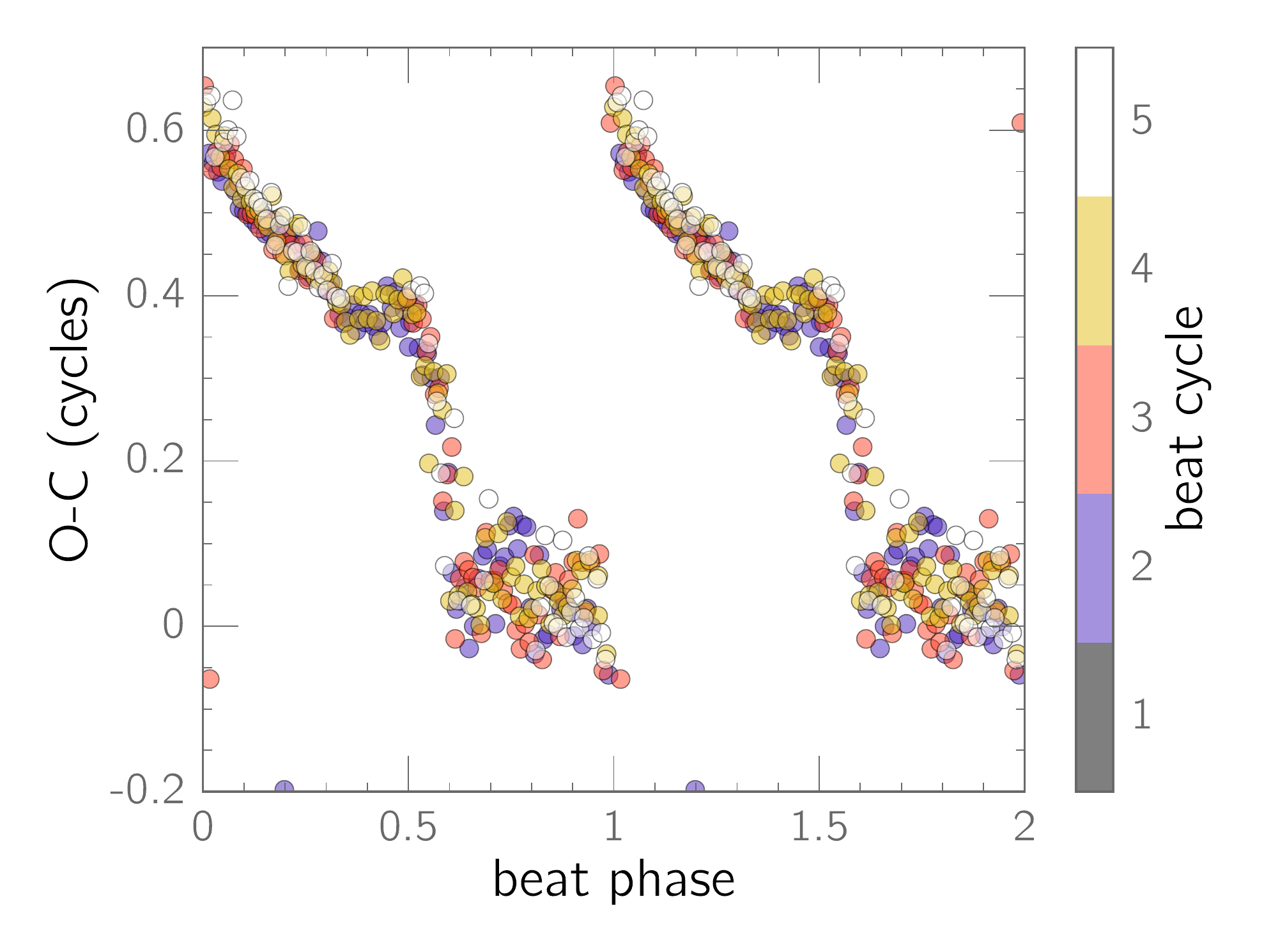}
    \caption{O-C diagram of the spin maxima with respect to our spin ephemeris. The upper-pole spin pulses are stable in phase, while the pulses from the lower pole show significant drift towards earlier phases, in apparent contravention of the prediction of \citet{gs97}. We discuss this tension in the text. Additionally, the drift of the lower pole's accretion region appears to cease at near the beginning of the depression in the light curve near $\phi_{beat}  = 0.3$.}
    \label{fig:O-C}
\end{figure}

\section{The spin period and its derivative} \label{sec:ephemeris}

\subsection{Spin ephemeris}

We extracted times of spin maxima by fitting a polynomial to each well-sampled photometric maximum. We visually inspected the fit quality of the polynomials, experimented with changing their degree, and found that fifth-order polynomials were the simplest polynomials that consistently offered satisfactory representations of the pulse maxima. Because the upper-pole pulses show the greatest phase stability (Fig.~\ref{fig:2d_lightcurves}), we use only these maxima to compute an ephemeris of $T_{max}[BJD] = 2458326.46492(17) + 0.0769522(11)\times E$. The duration of the \tess\ observation is insufficient to provide an independent estimate of the derivative of the spin period.

\subsection{Phase drift of the spin pulse}
\label{sec:phase_drift}

Fig.~\ref{fig:O-C} shows the residuals from our spin ephemeris. As expected from Fig.~\ref{fig:2d_lightcurves}, the lower pole's pulses experienced a significant drift across the beat cycle, while the upper pole's pulses were far more stable in phase. The residuals contain a pair of discontinuities, and we have defined beat phase 0.0 ($T_{0, beat}[BJD] = 2458328.9$) such that it coincides with the larger and sharper of the two. Interestingly, the phase drift of the lower pole ends at almost the exact same beat phase at which the depression in the beat-phased light curve (Fig.~\ref{fig:beat}) began.

It is noteworthy that the lower pole's pulses drifted towards earlier beat phases. \citet{gs97} presented a geometric model that predicted that the longitude of the accretion region on the WD should change across the beat cycle, as the accretion stream couples onto different field lines. According to the \citet{gs97} model, the spin maxima in a system in which $\omega > \Omega$ should show prograde motion, drifting towards later phases. 

There is mixed evidence for this effect in BY Cam and V1500 Cyg, the other two well-studied APs with $\omega > \Omega$. Fig.~4 in \citet{silber} shows a set of periodograms of segments of BY Cam's light curve that are sufficiently short that there are no internal pole-switches. In each of these segments, the measured period of the spin variability is shorter than the true spin period, which would mean that those spin pulses drifted towards earlier spin phases. Thus, the direction of the phase drift in BY Cam also contradicts the predictions of \citet{gs97}. However, unlike BY Cam, the direction of the phase drift in V1500 Cyg is prograde \citep[Fig.~6 in ][]{pavlenko}, consistent with \citet{gs97}. 

A possible explanation is that the \citet{gs97} model might need to take into account additional physical parameters in order to predict whether the drift of the pulses will be prograde or retrograde. For example any motion of the threading region would violate their assumption that the threading region's position is fixed in the binary rest frame. For example, there is evidence that the position of the threading region varies in V1432 Aql \citep{littlefield}. If the threading region were to undergo prograde motion across the beat cycle, it would change the threading radius---and, therefore, the magnetic colatitude of the accretion region. Moreover, the motion of the threading region would reduce the frequency at which different magnetic field lines would interact with the threading region, suppressing the motion predicted by \citet{gs97}. Likewise, the angle of deflection of the field lines at the threading region would likely change across the beat cycle and might contribute to the variable accretion geometry.

Doppler tomography of CD Ind's beat cycle, similar to the BY Cam study performed by \citet{schwarz}, would likely offer insight into whether the threading region drifts, helping to clarify the inconsistency between the \citet{gs97} model and the retrograde drift that results from our re-identification of the spin frequency. Moreover, detailed magneto-hydrodynamic modeling would be an excellent complement to the \citet{gs97} geometric model.

\subsection{Spin-period derivative}

If we have correctly re-identified the spin and orbital frequencies, the \citet{myers} spin ephemeris is actually an ephemeris of the $2\omega - \Omega$ sideband frequency. Nevertheless, the sideband ephemeris should still yield a reliable measurement of the spin-frequency derivative because its time derivative is twice that of the spin frequency (under the assumption of a negligible orbital-frequency derivative). Due to our reidentification of the frequencies, the \citet{myers} spin-frequency derivative was too large by a factor of 2, meaning that their resynchronization timescale of 6400 years would need to be doubled to 13000 years.

To test and refine the \citet{myers} $\dot{P}$, we combined our timings with the spin-pulse maxima from Table~4 in \citet{myers} and performed an O$-$C analysis on them, using the \citet{myers} ephemeris for the cycle count. Prior to conducting the O$-$C analysis, we carried out several checks to ensure compatibility between the two datasets. For instance, \citet{myers} only used timings of the lower pole, so we applied the same restriction to our data. Additionally, the \citet{myers} pulse maxima are expressed in HJD, so we converted these times to $BJD_{TDB}$ using routines in Astropy to enable a more direct comparison. We also excluded the first four pulse timings from \citet{myers} because of the nearly 11-year gap between those data and the rest of the CBA data.\footnote{\citet{myers} determined that the measured $\dot{P}$ changed only modestly if these four points were excluded.} The lack of a systematic time delay between the \tess\ and Myers light curves in Fig.~\ref{fig:overlapping_light_curve} indicates that the \tess\ pulse timings can be directly compared with ground-based measurements.

\begin{figure}
    \centering
    \includegraphics[width=\columnwidth]{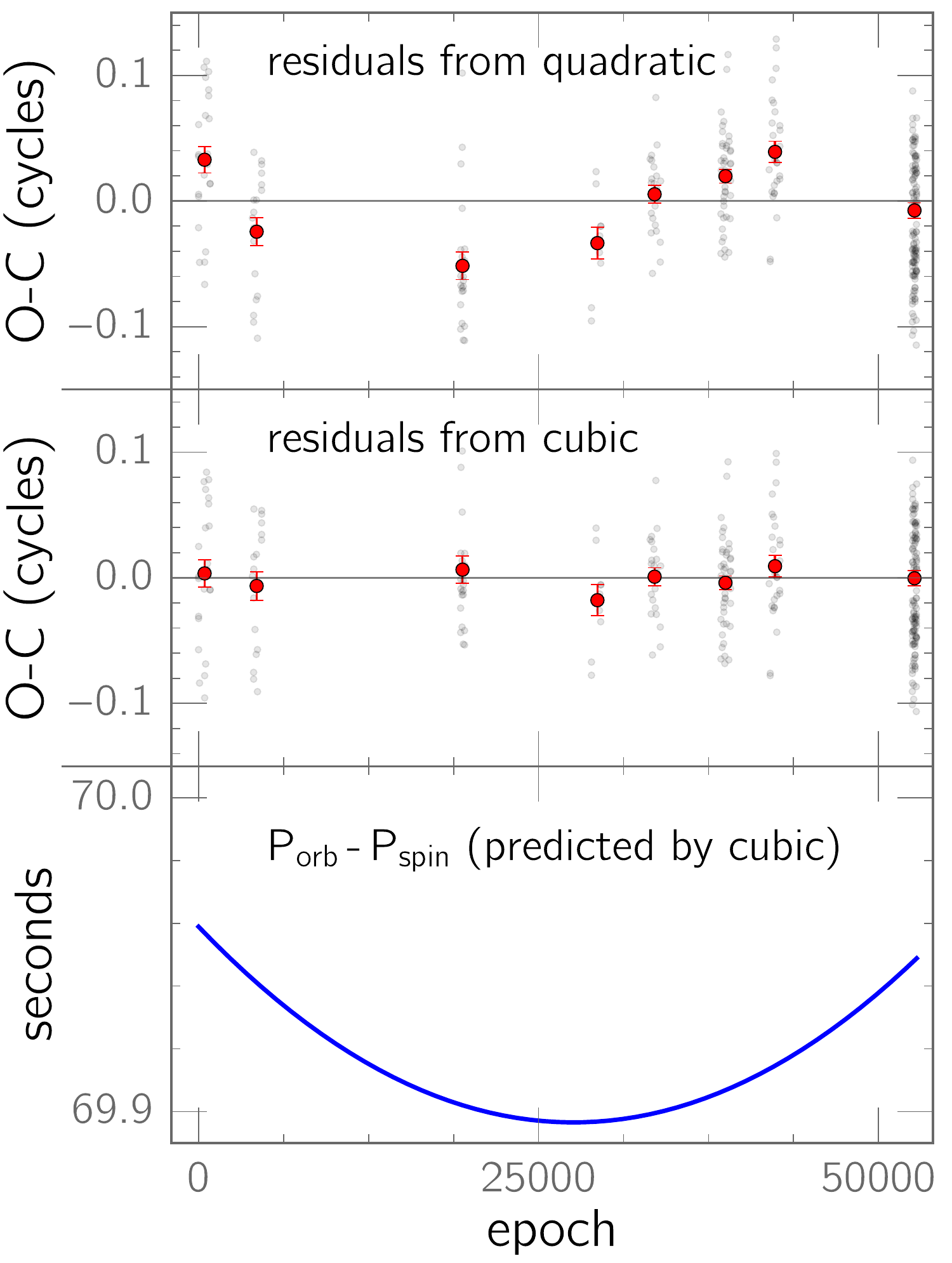}
    \caption{Top two panels: residuals from the best-fit quadratic and cubic ephemerides for the spin-pulse maxima of the lower pole, based on cycle counts from the \citet{myers} ephemeris. According to our new frequency identifications, the \citet{myers} ephemeris described the $2\omega - \Omega$ sideband, so the epoch would be the sideband cycle count. The far-right cluster of points is from the \tess\ light curve, while all other points are CBA measurements from \citet{myers}. The large red markers are yearly means. Bottom panel: the amount of asynchronism implied by the cubic ephemeris.\label{fig:ephemerides}}
\end{figure}

A second-order polynomial describes the CBA pulse timings well \citep{myers}, but the inclusion of the \tess\ timings requires the addition of a cubic term to the best-fit polynomial. Although this cubic ephemeris can nominally reconcile the pulse timings from \tess\ and \citet{myers}, it also predicts that the spin period of the WD began to increase during the CBA observations and that the system is currently evolving away from synchronization (Fig.~\ref{fig:ephemerides}).

We are skeptical of this interpretation. Such a result would contravene the basic theory of AP synchronization and would also imply that we are serendipitously observing CD Ind at an unusual point in its evolution, both of which are unlikely. A more plausible explanation is that on long timescales, the accretion geometry is not perfectly repeatable, causing the spin pulse to be an unreliable indicator of the underlying rotation of the WD. Any long-term variations in the accretion geometry (e.g., from a change in the mass-transfer rate) could masquerade as changes in the spin period of the WD, thwarting O$-$C analysis.

There are several observational grounds for questioning the long-term stability of the accretion geometry. \citet{myers} noted that over the ten-year baseline of CBA observations, the system occasionally dropped into a low-accretion state. Likewise, the waveform of the beat period in the \tess\ data (Fig.~\ref{fig:beat}) is starkly different than its counterpart in Fig.~6 of \citet{myers}. Indeed, \citet{myers} noted that the appearance of the beat cycle in their data varied from year-to-year for unknown reasons. In light of these observations, there is no guarantee that the accretion geometry remains stable on the long timescales needed to measure the evolution of the spin-period derivative.

While this scenario has implications for the \citet{myers} spin-down ephemeris, it is worth noting that their estimate of the resynchronization timescale accords with at least one theoretical prediction \citep[][p. 126]{campbell}, even after being doubled by our revised period identifications. It is possible that the geometry remained sufficiently stable during the \citet{myers} observations for the period derivative to be reliably measured.

\begin{figure}
    \centering
    \includegraphics[width = \columnwidth]{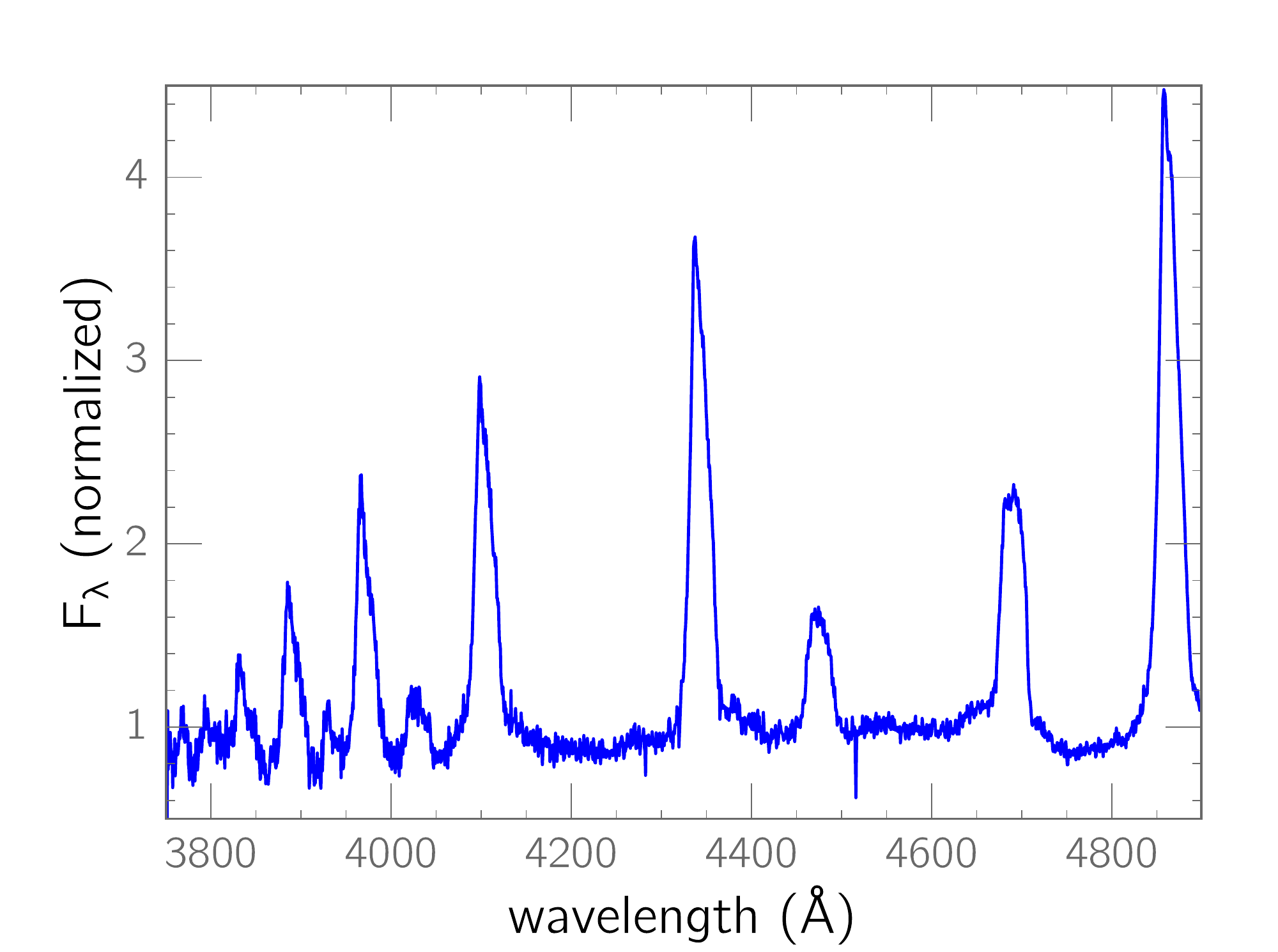}
    \caption{The continuum-normalized, average spectrum of CD Ind during the VLT observation on 2006 Aug. 13. The variable slit losses precluded a reliable flux calibration. The Balmer lines, He I $\lambda$ 4471 \AA, and He II $\lambda$ 4686 \AA\ are the most prominent lines. No velocity correction has been applied. }
    \label{fig:spectrum}
\end{figure}

\section{Spectroscopy}
\label{sec:spectra}

\begin{figure}
    \centering
    \includegraphics[width=\columnwidth]{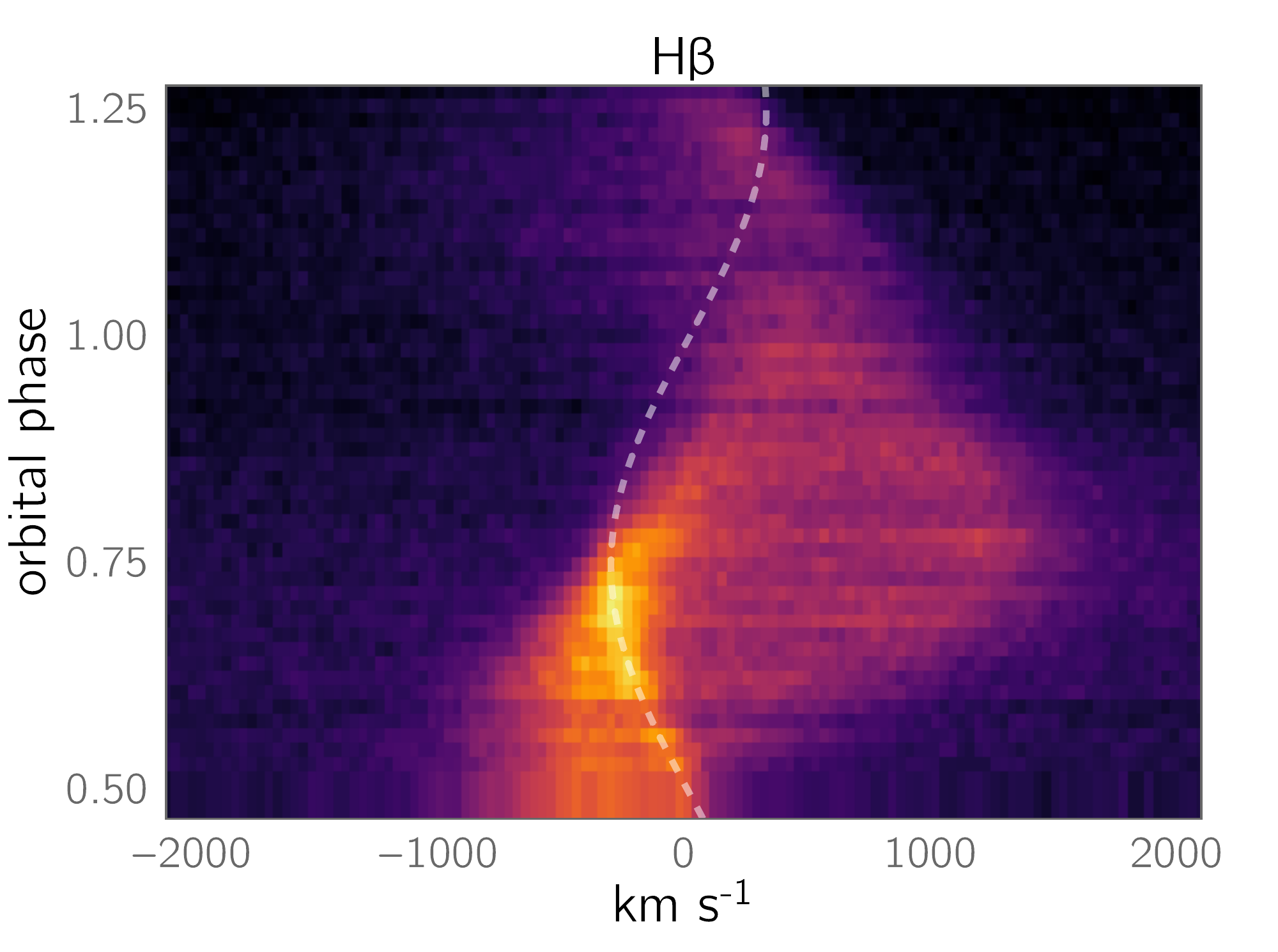}
    \caption{Trailed spectrum of H$\beta$. The dashed sinusoid is a visual fit to the narrow emission component visible between $0.5 \lesssim \phi_{orb} \lesssim 0.75$, which we attribute to the donor star's irradiated inner hemisphere. The fit parameters are $K_2 = 310$ km s$^{-1}$ and $\gamma = 20$ km s$^{-1}$, but the limited duration of the observations means that there is a considerable degeneracy between these two values. Nevertheless, we could not achieve a plausible-looking fit with phase shifts larger than $\pm$0.02, so we are confident that the calculated orbital phases are accurate both in this figure and in the tomograms. }
    \label{fig:trailed_spec}
\end{figure}

\begin{figure*}
\gridline{
\fig{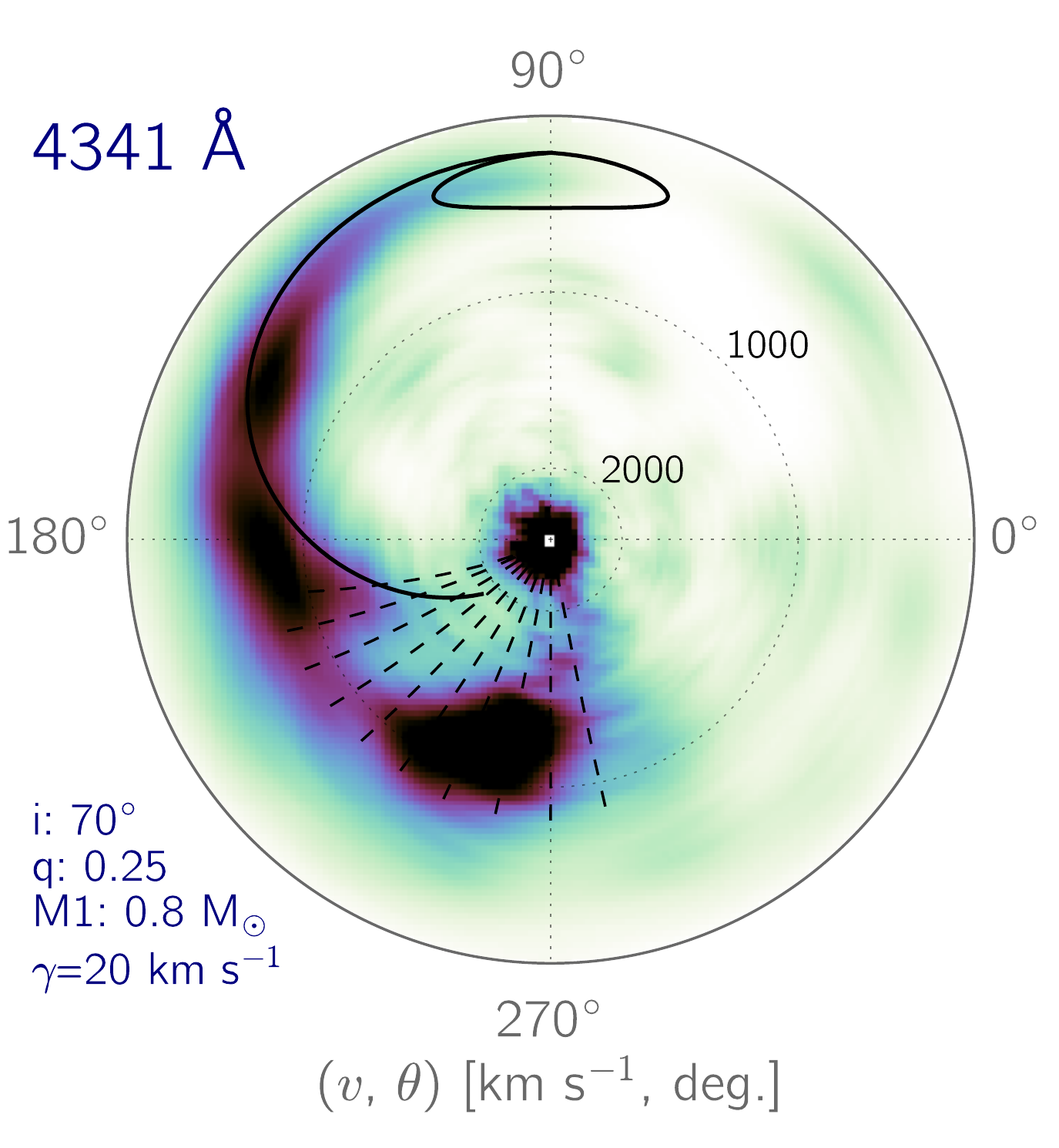}{0.25\textwidth}{}
\fig{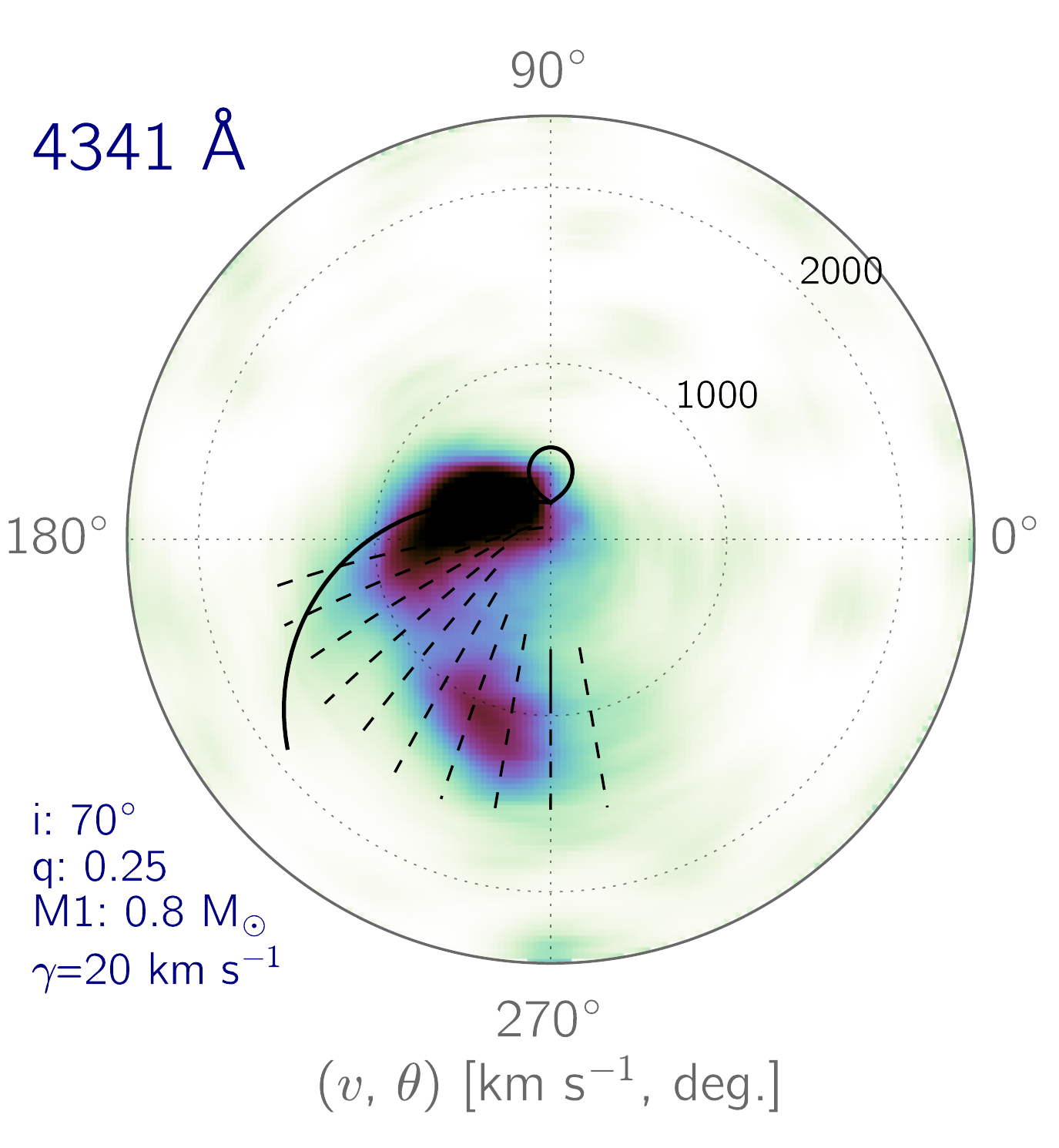}{0.25\textwidth}{}
\fig{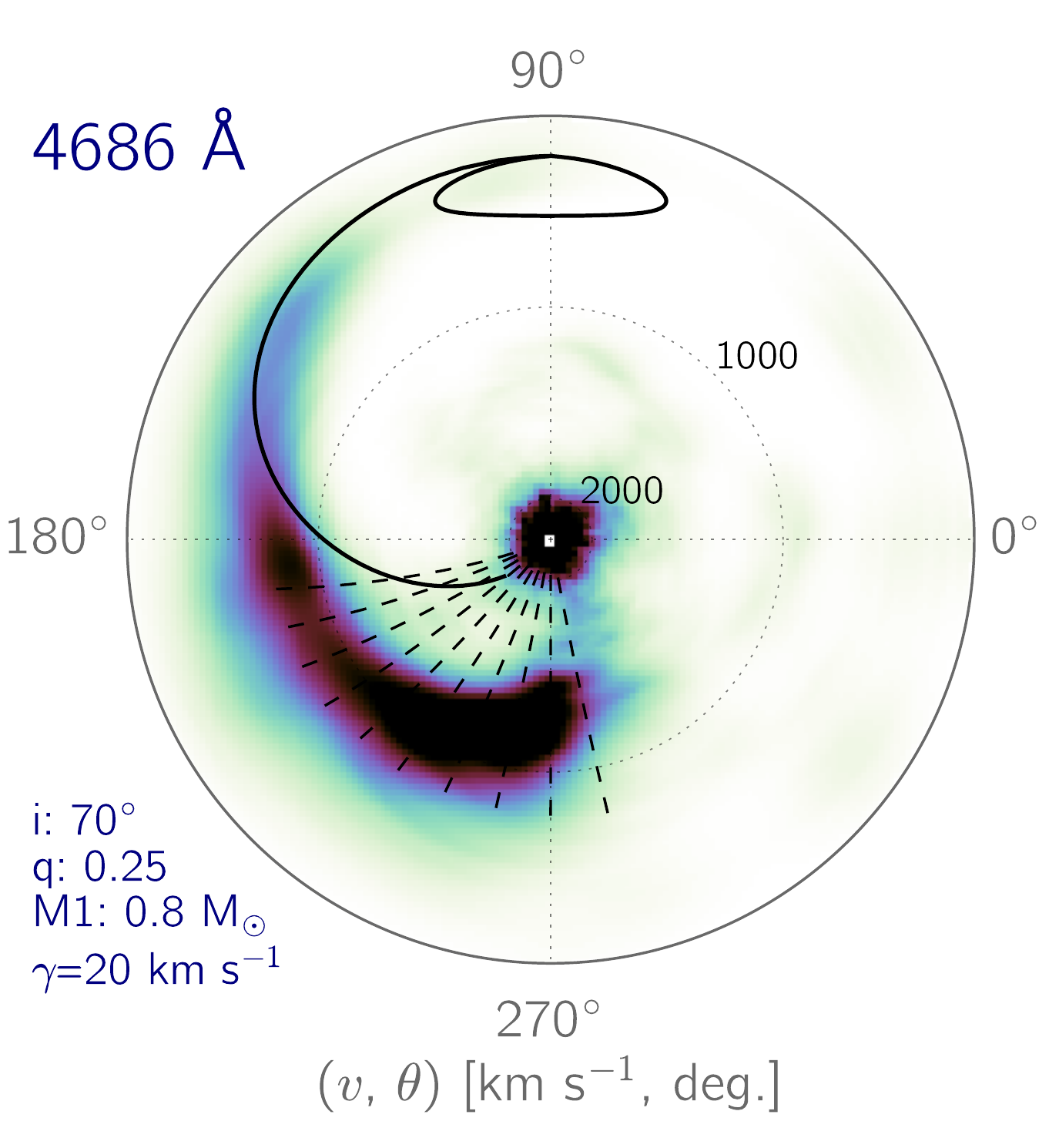}{0.25\textwidth}{}
\fig{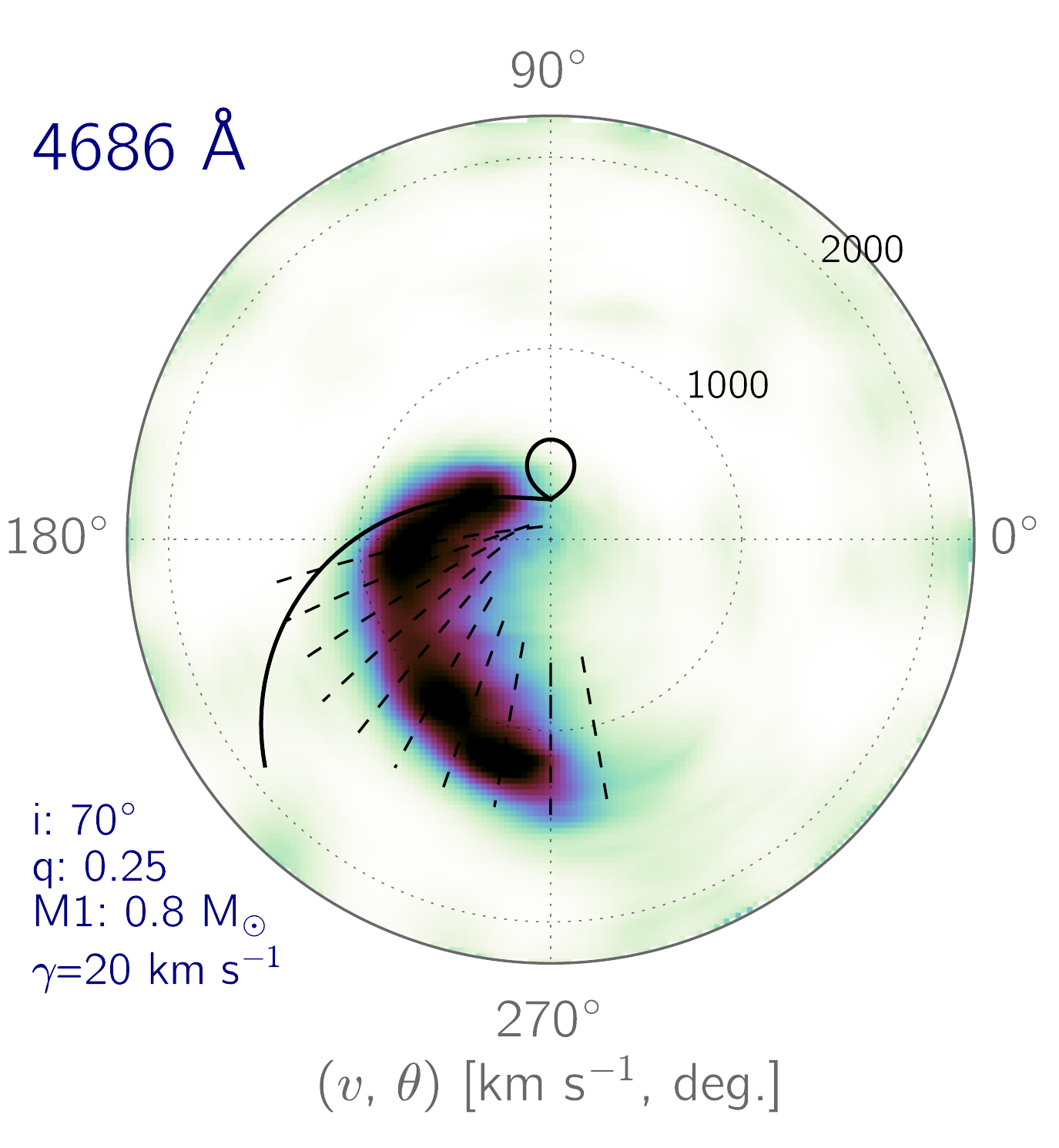}{0.25\textwidth}{}
}
\caption{Inside-out and standard Doppler tomograms of H$\gamma$ (left two panels) and He II $\lambda$ 4686\AA\ (right two panels) from spectra obtained in 2006. We have assumed typical values for the mass ratio $q$ and the WD mass $M_1$, and we further assume that the orbital inclination $i$ is relatively high in accordance with \citet{ramsay00}. These three values affect only the velocity overlay (black lines), not the actual computation of the tomograms. The thick black line indicates the velocity coordinates of the secondary and the ballistic accretion stream, while the dashed black lines represent various magnetic trajectories of material that becomes magnetically confined at different points along the stream. 
\label{fig:tomograms}}
\end{figure*}

The average spectrum of CD Ind from the 2006 observation (Fig.~\ref{fig:spectrum}) is typical for a polar, with strong emission at the Balmer lines and He II $\lambda$ 4686 \AA. The He I lines at $\lambda$ 4023 \AA\ and 4471 \AA\ are also present, the latter being significantly more prominent than the former. The lack of additional high-excitation lines, such as the CIII/NIII Bowen blend, suggests that CD Ind was in a relatively low accretion state during the observations.

Utilizing the code described in \citet{kotze}, we mapped the spectra as Doppler tomograms using both standard and inside-out projections. The Balmer lines in the spectra contain an emission component from the irradiated inner hemisphere of the secondary (Fig.~\ref{fig:trailed_spec}), and by modeling this emission component with a sine wave, we estimate the blue-to-red crossing to have occurred at BJD = 2453960.9025(16). We use this information to phase the spectra to the binary orbit.

In Fig.~\ref{fig:tomograms}, we display tomograms of two representative lines (H$\gamma$ and He II $\lambda$4686 \AA). Both lines show a well-defined ballistic accretion stream, though it stands out with greater contrast in H$\gamma$. The magnetically confined portion of the flow is prominent, particularly in the high-excitation He II $\lambda$4686 \AA\ line. The inside-out projection shows high-velocity radial inflow with a direction of just over 270$^{\circ}$ in velocity space, which is consistent with an accretion region that leads the secondary by $\sim90^{\circ}$ in physical coordinates.\footnote{See Fig.~1 in \citet{kotze} for a schematic relating physical space with velocity space.} There is also high-velocity emission near the origin of the inside-out plots, but we are reluctant to ascribe an astrophysical significance to it, particularly in light of the highly variable slit losses during the observations. Moreover, there is no obvious counterpart to this emission in the trailed spectrum in Fig.~\ref{fig:trailed_spec}, suggesting that it is an artifact.

Emission from the donor star is not apparent in the tomograms, likely because of the combination of incomplete orbital coverage during the original spectroscopic observations and geometric occultation of the line-forming region on the inner hemisphere. In total, only $\sim$one-third of our spectra show emission from the inner hemisphere, and it is questionable whether Doppler tomography could reliably recover a feature present in significantly less than half of the input spectra.

The Doppler tomograms of CD Ind are remarkable in comparison to other asynchronous polars. Doppler tomography of BY Cam \citep{schwarz}, IGR J19552+0044 \citep{tov17}, and V1432 Aql \citep[][Fig. 10]{schwope01} provides compelling evidence that the accretion flows in those systems are far more extended than in synchronous polars. Specifically, their accretion curtains subtend a wide azimuthal extent and are present at azimuths not normally observed in synchronous polars. For example, in BY Cam, \citet{schwarz} estimated that the stream must travel at least halfway around the WD in order to produce the extended accretion curtain in their observations. In contrast, the emission in CD Ind is confined primarily to the left half of each tomogram, and while there is evidence of a bright, extended magnetic flow, the azimuthal extent of the highest-velocity emission is relatively narrow. While this suggests a wider accretion curtain than is normally seen in synchronous polars, it was nevertheless substantially smaller than the curtains in the other three systems. Without better spectroscopic coverage of CD Ind's beat period, it is impossible to know whether these tomograms are representative of the system's typical behavior.

The duration of the spectroscopic observations is too short to enable us to obtain a precise measurement of the period of the emission from the secondary, but should a future study obtain a high-precision measurement of this period, it will provide a stringent test of our re-identification of $\omega$ and $\Omega$.

\begin{figure}
    \centering
    \includegraphics[width=\columnwidth]{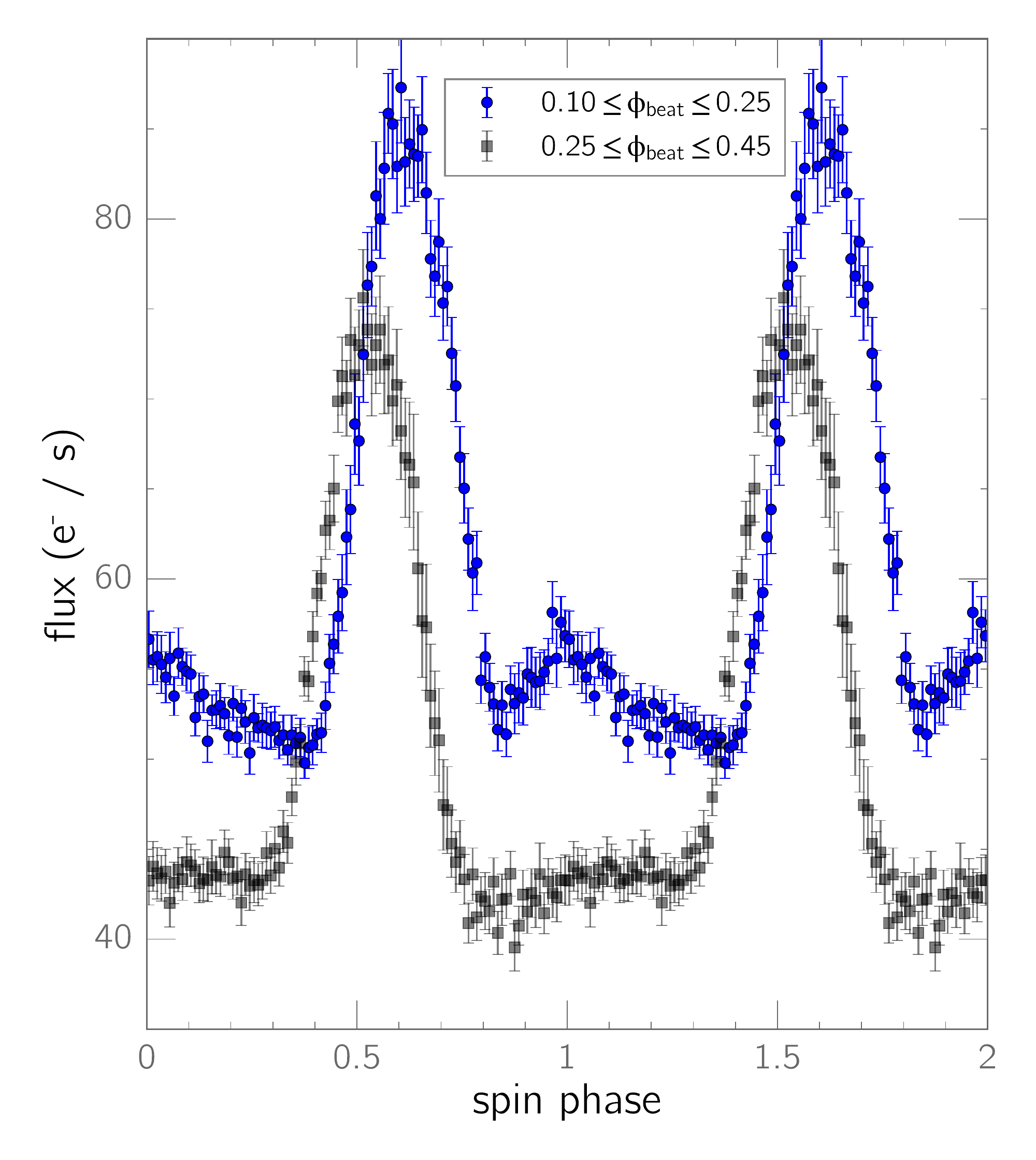}
    \caption{Phase-averaged profiles of the lower pole's spin pulse before and during the depression seen in Fig.~\ref{fig:beat}. Before the start of the depression, there was a weak hump near the spin phase of the upper pole's pulse; during the depression, this hump disappeared, and the light curve faded at all spin phases. We hypothesize that when the accretion switched to the lower pole, there was still residual accretion onto the upper pole and that the cessation of this accretion resulted in the depression in the light curve of the beat cycle.}
    \label{fig:lower_pole}
\end{figure}

\begin{figure*}
    \centering
    \includegraphics[width=\textwidth]{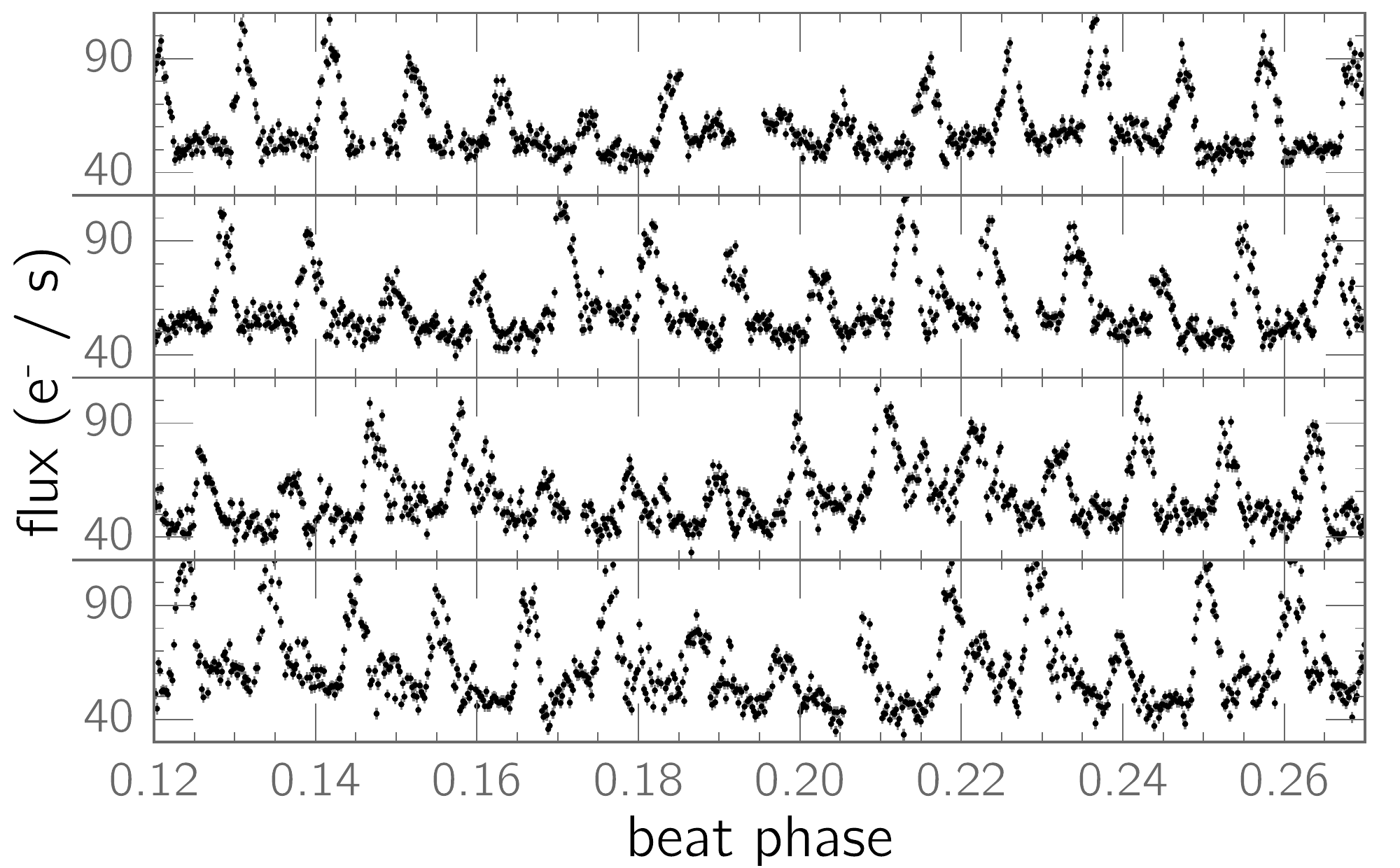}
    \caption{Variability of the photometric maximum from the lower pole near $\phi_{beat}\sim$0.2. Each panel shows a different beat cycle, and they are in chronological order from top to bottom. During the first beat cycle, the pulse amplitude dropped near $\phi_{beat}\sim$0.2 While this behavior was not apparent in the second beat cycle, the pulse amplitude again dropped near $\phi_{beat}\sim$0.2 in the final two beat cycles, and the pulse waveform briefly bifurcated. The tendency of these variations to recur near $\phi_{beat}\sim$0.2 suggests that the accretion geometry is somewhat unstable near this particular beat phase.}
    \label{fig:geometry_stability}
\end{figure*}

\section{Discussion}

\subsection{Nature of the dip in the beat cycle}

As Fig.~\ref{fig:beat} establishes, each beat cycle contained a depression $\sim$20\% deep lasting for just under a quarter of the beat cycle. When the depression began, the phase drift of the lower pole ceased, and the depression ended abruptly with the switch from the lower pole to the upper pole.

While the nature of this dip is ambiguous, we propose that it might be the consequence of the cessation of residual accretion onto the upper pole during the half of the beat cycle in which the lower pole is active. As shown in Fig.~\ref{fig:lower_pole}, after the accretion flow switched to the lower pole, there was evidence of a weak hump near the expected phase of the upper pole. This hump disappeared at the same beat phase at which the depression began. If there were residual accretion onto the upper pole, it would provide an extra source of flux to offset the disappearance of the lower accretion region behind the limb of the WD. At some beat phase, the upper pole would finally become so energetically disfavored that it would no longer accrete significantly, and the light curve would fade across all spin phases, resulting in the depression. Later in the beat cycle, the resumption of accretion at the upper pole would end the depression. 

If our proposed explanation is correct, then the intervals between pulses during the depression would be the only times during which no accretion region was visible. Although we are not aware of any contemporaneous X-ray or polarimetric observations of CD Ind during or near the \tess\ observation, such observations could test this prediction.

\subsection{Stability of the accretion geometry}

The \tess\ dataset presents an unprecedented opportunity to assess the stability of the accretion geometry across consecutive beat cycles. One way of measuring the repeatability of the accretion geometry is to examine the O$-$C diagram for evidence that the pulse timings from one beat cycle differed from those in another. To that end, the markers in the O$-$C values in Fig.~\ref{fig:O-C} are color-coded to indicate the beat cycle during which they were obtained. The consistency of the O$-$C across all beat cycles attests to the overall repeatability of the accretion geometry and its complex variations across the beat cycle.

Nevertheless, a visual inspection of the light curve reveals that there were some variations between beat cycles, the most conspicuous of which manifested itself as a brief change in the waveform of the lower-pole spin pulse near $\phi_{beat}\sim0.2$ (Fig.~\ref{fig:geometry_stability}). In three of the four beat cycles that covered this beat phase, the pulse's amplitude dropped, and its width increased. While this was happening, the pulse occasionally split into two weak components. The timescale for these variations was only 3-5 orbits.

The \tess\ observations do not directly reveal the stability of the accretion geometry on timescales of years, but the disagreement between the \tess\ and \citet{myers} pulse timings provides indirect evidence that the repeatability of CD Ind's accretion geometry might vary on timescales of years. If this is so, it would distinguish CD Ind from V1432 Aql, a system whose eclipse timings exhibit accretion-geometry-induced variations that are stable on timescales of $\sim$2 decades \citep{littlefield}. Unlike CD Ind, V1432 Aql has always been observed in a state of high mass-transfer, a difference that likely accounts for the excellent repeatability of its accretion geometry. Since the mass-transfer rate is one of the parameters that determines where the accretion flow is threaded, the low states noted in CD Ind by \citet{myers}  mean that its accretion geometry is probably less stable than that of V1432 Aql on long timescales. Thus, while it is possible that an AP's accretion geometry can remain consistent across years-long timespans, this must be established individually for each system.

\subsection{Photometric contribution of the secondary}

When interpreting the light curves, it is helpful to know whether the secondary contributes significantly. The spectral type of CD Ind's secondary is unknown, but an M6V star would be typical for CD Ind's orbital period \citep[][Fig. 15]{knigge}. A non-irradiated, isolated M6 dwarf would have an absolute SDSS $i'$ magnitude of $12.95\pm0.25$ \citep{west}. CD Ind's Gaia distance of $244\pm6$~pc is equivalent to a distance modulus of 6.9~mag. Thus, at CD Ind's distance, an M6 dwarf would have an expected $i'$ magnitude of $19.9\pm0.3$ if extinction were neglected. The expected contribution of the WD would be negligible. The ninth data release of APASS \citep{apass} measured CD Ind to be $i' = 15.44$ (with no uncertainty specified), and we presume that the luminosity excess above the stellar photospheres is due to accretion. If the APASS magnitude is typical of CD Ind, the companion would contribute an almost negligible $\lesssim$2\% of the flux in that bandpass.

The system's cyclotron flux can induce 1.4-magnitude cyclotron variations in the nearby $I$ band \citep{schwope}, and since we do not know whether the APASS magnitude was obtained at the peak of the cyclotron flux, it is also instructive to estimate the contamination by the secondary if CD Ind is typically fainter in $i'$ than its APASS magnitude would indicate. If CD Ind faded by 1.4 mag to $i' = 16.9$ in the absence of cyclotron radiation, the relative $i'$-band contribution of the M6V companion would increase to just $\sim6$\%. 

While these calculations could be improved by knowing both the secondary's spectral type and how irradiated it is, they nevertheless establish that the secondary is unlikely to be a major contributor to the \tess\ light curve.

\section{Conclusion}

We have argued that the long-accepted spin frequency in CD Ind is actually the $2\omega - \Omega$ sideband and that the long-accepted orbital frequency is the true spin frequency. These revised frequency identifications would require the \citet{myers} spin-period derivative to be halved and the associated synchronization timescale to be doubled to 13000 years. When phase-folded on the newly identified spin frequency, the \tess\ light curve suggests an accretion geometry consisting of two regions on opposite hemispheres of the WD. 

The waveforms of the two accretion regions imply that one was almost always visible during the \tess\ observation, while the other was hidden behind the limb of the WD for over half of the WD's rotational period. The accretion flow alternated between the two poles twice per beat cycle, and the process of pole-switching occurred in $\sim$10 binary orbits.

Regardless of the correct period identification, the \tess\ timings are inconsistent with the ephemeris reported by \citet{myers}. We speculate that the spin pulse might be an unreliable indicator of the WD's rotational period across very long periods of time, but this does not necessarily invalidate their measurement of $\dot{P}$ on shorter timescales.

Doppler tomography of CD Ind from 2006 reveals that the magnetically confined portion of the accretion curtain was unusually bright and that the accreting pole led the secondary star by $\sim90^{\circ}$, more than what is normally observed in synchronous polars. Nevertheless, the tomograms obtained at this particular beat phase show a less extended accretion flow than in Doppler tomography of other asynchronous polars. 

It should be possible to obtain ironclad identifications of the spin and orbital frequencies by measuring the spectroscopic period of the secondary, whose irradiated inner hemisphere is apparent in our trailed spectra. This measurement would yield an unambiguous measurement of the orbital period which, in conjunction with the known length of the beat period, would offer a direct test of our proposed identification of the spin period.\\

\section*{Acknowledgments}

We thank the referee for a careful reading of this manuscript and for a number of helpful suggestions.

We acknowledge useful discussions with Pasi Hakala, Gavin Ramsay, and David Buckley.

We thank Joe Patterson for his comments about an earlier version of this manuscript. 

P.S. acknowledges support from NSF grant AST-1514737.

M.R.K. is funded by a Newton International Fellowship provided by the Royal Society.

\software{Astropy \citep{astropy13, astropy18}, {\tt lightkurve} \citep{lightkurve}}

\end{document}